\documentclass[aps,prd,nofootinbib,twocolumn,superscriptaddress,preprintnumbers,balancelastpage,longbibliography]{revtex4-1}

\usepackage[utf8]{inputenc}
\usepackage{amsmath,amssymb,mathtools,bm}
\usepackage{graphicx, color, hepunits}
\usepackage[dvipsnames]{xcolor}
\usepackage{float}
\usepackage{multirow}
\usepackage{braket}
\usepackage{hyperref} 
\usepackage{float}

\usepackage{caption}
\usepackage[lofdepth,lotdepth,justification=RaggedRight]{caption}

\usepackage{subcaption}

\hypersetup{
    colorlinks=true,      
    linkcolor=blue,        
    citecolor=blue,        
    filecolor=magenta,     
    urlcolor=blue          
}
\usepackage[utf8]{inputenc}
\usepackage[english]{babel}
\usepackage{lipsum}
\usepackage{mathrsfs}

\usepackage{tensor}

\def\beq{\begin{equation}}
\def\eeq{\end{equation}}

\def\Beq{\begin{equation}\begin{aligned}}
\def\Eeq{\end{aligned}\end{equation}}

\def\bea{\begin{eqnarray}}
\def\eea{\end{eqnarray}}

\def\beq{\begin{equation}}
\def\eeq{\end{equation}}
\def\bea{\begin{eqnarray}}
\def\eea{\end{eqnarray}}

\DeclareMathOperator{\sech}{sech}

\begin{document}

\title{The Electroweak Sphaleron Revisited: \\
II. Study of Decay Dynamics}

\author{Konstantin T. Matchev}
\affiliation{Department of Physics and Astronomy, University of Alabama, Tuscaloosa, AL 35487, USA}

\author{Sarunas Verner}
\affiliation{Institute for Fundamental Theory, Physics Department, University of Florida, Gainesville, FL 32611, USA}

\vspace{0.5cm}

\date{\today}

\begin{abstract}
We present a comprehensive analysis of electroweak sphaleron decay dynamics, employing both analytical techniques and high-resolution numerical simulations. Using a spherically symmetric ansatz, we reformulate the system as a $(1+1)$-dimensional problem and analyze its stability properties with current Standard Model parameters ($m_H = 125.1$ GeV, $m_W = 80.4$ GeV). We identify precisely one unstable mode with eigenvalue $\omega_{-}^2 \simeq -2.7m_W^2$ and numerically evolve the full non-linear field equations under various initial conditions. Through spectral decomposition, we quantify the particle production resulting from the sphaleron decay. Our results demonstrate that the decay process is dominated by transverse gauge bosons, which constitute approximately 80\% of the total energy and multiplicity, while Higgs bosons account for only 7-8\%. On average, the sphaleron decays into 49 $W$ bosons and 4 Higgs bosons. The particle spectra consistently peak at momenta $k \sim 1 - 1.5 \, m_W$, reflecting the characteristic size of the sphaleron. Remarkably, these properties remain robust across different decay scenarios, suggesting that the fundamental structure of the sphaleron, rather than specific triggering mechanisms, determines the decay outcomes. These findings provide distinctive experimental signatures of non-perturbative topological transitions in the electroweak theory, with significant implications for baryon number violation in the early universe and potentially for high-energy collider physics.
\end{abstract}

\maketitle

\section{Introduction}
\label{sec:intro}
The electroweak sphaleron—a topologically non-trivial, unstable solution to the classical field equations of the Standard Model (SM)—plays a central role in non-perturbative phenomena ranging from baryon number violation to early universe baryogenesis. This saddle-point configuration sits at the energy barrier separating vacua with different Chern-Simons numbers and mediates transitions that violate baryon (B) and lepton (L) number through the chiral anomaly~\cite{Adler:1969gk, Bell:1969ts}. Initially identified by Klinkhamer and Manton~\cite{Klinkhamer:1984di}, the sphaleron has become central to our understanding of topological transitions in gauge theories and their physical consequences. The theoretical framework for baryon number violation in the electroweak theory was first established by 't Hooft~\cite{tHooft:1976rip, tHooft:1976snw}, who showed that instanton solutions of the $\mathrm{SU}(2)$ gauge theory could induce $B$ and $L$ violating interactions. At zero temperature, such processes are exponentially suppressed by a factor of $\exp(-4\pi/\alpha_W) \sim 10^{-161}$, effectively rendering them unobservable. However, a crucial insight came from Kuzmin, Rubakov, and Shaposhnikov~\cite{Kuzmin:1985mm}, who determined that at high temperatures in the early universe, these baryon-number violating transitions would occur at unsuppressed rates. This realization established the electroweak sphaleron as a key element in early universe baryogenesis scenarios.

The sphaleron solution was identified in the limit where the weak mixing angle $\theta_W$ vanishes, which simplifies the theory to an $\mathrm{SU}(2)_{\mathrm{W}}$ gauge theory coupled to a scalar Higgs doublet. While this approximation decouples the $\mathrm{U}(1)_{\mathrm{Y}}$ hypercharge, it remains remarkably effective since the energy functional varies only slightly with $\theta_W$~\cite{Manton:1983nd, Klinkhamer:1990fi, Kunz:1992uh, DOnofrio:2014rug}. The electroweak vacuum exhibits topological degeneracy characterized by integer values of the Chern-Simons number $N_{\mathrm{CS}}$. In this context, the sphaleron configuration corresponds to $N_{\mathrm{CS}}=1/2$, positioning it at the midpoint between neighboring vacua with $N_{\mathrm{CS}}=0$ and $N_{\mathrm{CS}}=1$~\cite{Manton:2004tk}.

Recent years have witnessed renewed interest in sphaleron physics, particularly regarding potential manifestations in high-energy collider experiments. Several authors have investigated whether sphaleron-mediated baryon and lepton number violation might be observable at current or future accelerators~\cite{Gibbs:1994cw, Ringwald:2002sw, Bezrukov:2003er, Bezrukov:2003qm}. Of particular note is the work by Tye and Wong~\cite{Tye:2015tva}, who developed a Bloch-wave description of the periodic vacuum structure, suggesting that above a critical energy threshold ($E \gtrsim E_{\mathrm{sph}}$), baryon number violation could proceed without exponential suppression. This theoretical framework has been extended to include realistic collider observables~\cite{Ellis:2016dgb, Ellis:2016ast}, and has even prompted the CMS collaboration to conduct dedicated searches for anomalous multi-fermion events consistent with sphaleron-induced transitions~\cite{CMS:2018ozv}. Despite this renewed interest, the quantitative characterization of the sphaleron has seen relatively few updates since the initial studies in the 1980s. Most earlier analyses preceded the Higgs boson discovery and frequently assumed approximate mass equivalence between the Higgs and W bosons ($m_H \simeq m_W$) for computational simplicity. Now that we have measured the Higgs mass to be $m_H \simeq 125.1$ GeV, significantly larger than $m_W \simeq 80.4$ GeV~\cite{ParticleDataGroup:2024cfk}, it is important to revisit the sphaleron with physically accurate SM parameters.

The importance of the sphaleron extends well beyond high-energy physics into cosmology. By violating $(B+L)$ while conserving $(B-L)$, sphaleron processes can transform a primordial lepton asymmetry into the observed baryon asymmetry, forming the foundation for leptogenesis models~\cite{Fukugita:1986hr, Davidson:2008bu}. Current lattice simulations with realistic Higgs mass indicate that the electroweak phase transition is a smooth crossover occurring at $T_c \simeq 159$ GeV~\cite{DOnofrio:2014rug}. Above this temperature, $B+L$ violating processes are rapid, while below it, the sphaleron rate decreases substantially, with an estimated freeze-out temperature of $T_* \simeq 132$ GeV. This temperature threshold plays a critical role in constraining viable baryogenesis mechanisms~\cite{Cohen:1993nk, Rubakov:1996vz, Trodden:1998ym, Morrissey:2012db, Shaposhnikov:1987tw}.

In this paper, we investigate the dynamical process of sphaleron decay in the $\mathrm{SU}(2)$-Higgs theory. While the properties of the static sphaleron configuration have been extensively studied, the real-time evolution as the sphaleron decays toward the vacuum state remains less thoroughly explored~(see Ref.~\cite{Hellmund:1991ub} for some earlier work). We employ both analytical methods and high-resolution numerical simulations to track this topological transition and characterize the resulting particle production. Our approach begins with the classical field equations derived from the $\mathrm{SU}(2)$-Higgs Lagrangian. Using a spherically symmetric ansatz, we reduce the system to a $(1+1)$-dimensional problem that remains computationally tractable while preserving the essential topological features. Working with current SM parameters, we first compute the static sphaleron solution and analyze its stability, confirming the presence of exactly one unstable mode with eigenvalue $\omega_{-}^2 = -2.7m_W^2$~\cite{MatchevVerner}. This negative eigenvalue defines the preferred direction for sphaleron decay, which we then explore through numerical simulations.

The main focus of our work is to characterize the particle production resulting from the sphaleron decay. Through spectral analysis of the asymptotic field oscillations, we determine the energy and multiplicity distributions of gauge and Higgs bosons produced during this non-perturbative process. Our calculations reveal that the sphaleron typically decays into approximately 48 gauge bosons and 4 Higgs bosons, with transverse gauge polarizations dominating the spectrum. The particle momentum distributions consistently peak around $k \sim 1$-$1.5\,m_W$, reflecting the characteristic spatial scale of the sphaleron. Notably, these spectral properties show minimal variation across different decay scenarios, suggesting they represent intrinsic signatures of sphaleron-mediated processes rather than artifacts of particular initial conditions.

The paper is organized as follows. In Section~\ref{sec:fieldequations}, we introduce the $\mathrm{SU}(2)$-Higgs theory, implement the spherically symmetric ansatz, and derive the resulting field equations. Section~\ref{sec:fieldequationsminimum} analyzes these equations around the vacuum state, developing a systematic framework for the spectral decomposition of small-amplitude oscillations. In Section~\ref{sec:energyanalysis}, we derive the energy functional and establish the connection between classical field amplitudes and the particle content. Section~\ref{sec:numerical} presents our numerical simulations of sphaleron decay under four distinct initial conditions, providing a comparative analysis of the field evolution and resulting particle spectra. Finally, in Section~\ref{sec:conclusions}, we summarize our results.
\section{\boldmath $SU(2)$ Classical Field Equations, Spherical Symmetry, and Sphaleron Solution}
\label{sec:fieldequations}
Our analysis of sphaleron decay begins with the $SU(2)$ Yang-Mills theory, which captures the essential non-perturbative dynamics of the electroweak sector in the limit where the weak mixing angle vanishes ($\theta_W \rightarrow 0$).\footnote{The validity of this simplification has been rigorously established through several independent analyses~\cite{Manton:1983nd, Klinkhamer:1990fi, Kunz:1992uh}. These studies demonstrate that incorporating the physical value $\sin^2\theta_W \simeq 0.23$ modifies the sphaleron energy by only $\simeq 1\%$. Consequently, our adopted value $E_{\rm{sph}} \simeq 9.0~\rm{TeV}$ provides sufficient precision for both theoretical investigations and phenomenological applications concerning baryon and lepton number violation in the early universe and potentially at future colliders.} The dynamics of this system are governed by the Lagrangian density:
\begin{equation}
    \label{eq:SU2lagrangian}
    \mathcal{L} \; = \; -\frac{1}{4}(W_{\mu\nu}^a)^2 + |D_\mu\Phi|^2 - \lambda\left(|\Phi|^2 - \frac{v^2}{2}\right)^2 \, ,
\end{equation}
where $W_{\mu\nu}^a = \partial_\mu W_\nu^a - \partial_\nu W_\mu^a + g\epsilon^{abc}W_\mu^b W_\nu^c$ is the field strength tensor for the $SU(2)$ gauge field $W_\mu^a$, and $D_\mu\Phi = \partial_\mu\Phi - \frac{i}{2}gW_\mu^a\sigma^a\Phi$ is the covariant derivative acting on the Higgs doublet $\Phi$. Here, $g$ is the $SU(2)$ gauge coupling constant, $\sigma^a$ are the Pauli matrices that generate the $SU(2)$ Lie algebra, $\lambda$ is the Higgs self-coupling, and $v$ is the Higgs vacuum expectation value.

In this section, we first introduce the spherically symmetric ansatz for the gauge and Higgs fields and derive the resulting field equations. We then examine the static sphaleron solution and analyze its stability properties, with particular focus on the unstable mode that governs its decay dynamics.

\subsection{Spherically Symmetric Ansatz}
\label{subsec:sphericalansatz}
A key challenge in analyzing the sphaleron lies in managing the mathematical complexity of the full $\mathrm{SU}(2)$-Higgs theory. To address this, we implement a spherically symmetric ansatz~\cite{Ratra:1987dp, Akiba:1988ay}, which transforms the three-dimensional field theory into a $1+1$-dimensional system. This dimensional reduction significantly simplifies the computational approach while preserving the essential topological properties, particularly the Chern-Simons winding number that characterizes the distinct vacuum states:
\begin{align}
\label{eq:sphans1}
W_0^a(x) \; &=\; \frac{1}{g}\,G(r,t)\,\frac{x_a}{r}\,,
\\
\label{eq:sphans2}
    W_j^a(x)  &\; = \;\frac{1}{g} \left[\frac{(f_A(r,t) - 1)}{r^2} \varepsilon_{jam} x_m \right.~\nonumber \\ 
    &\left.  + \frac{f_B(r,t)}{r^3}\left(r^2 \delta_{ja} - x_j x_a \right) + \frac{f_C(r,t)}{r^2}x_jx_a \right] \, ,
\\
\label{eq:sphans3}
\Phi(x) \; &=\; \frac{v}{\sqrt{2}}\!
\left[\,H(r,t)
   + i\,K(r,t)\,\frac{\boldsymbol{\sigma}\!\cdot\!\mathbf{x}}{r}
\right]
\begin{pmatrix}
 0 \\ 1
\end{pmatrix} \, .
\end{align}

In these expressions, we denote the magnitude of the position vector as $r = |\mathbf{x}|$ and employ the projection of Pauli matrices along the radial direction via $\boldsymbol{\sigma} \cdot \mathbf{x} = \sigma^a x_a$. The time-dependent dynamics of the system is entirely encoded in six radial profile functions: $G(r,t)$ for the temporal gauge component, $f_A(r,t)$, $f_B(r,t)$, and $f_C(r,t)$ for the spatial gauge components, and $H(r,t)$ and $K(r,t)$ for the Higgs field. The structure of this parametrization naturally embeds the hedgehog topology, where the isospin orientation aligns with the spatial direction, that characterizes non-trivial gauge field configurations with non-zero winding number, making it ideally suited for analyzing the sphaleron and its decay.

Applying the principle of least action to the Lagrangian \eqref{eq:SU2lagrangian} yields the coupled field equations
\begin{align}
    D_{\mu} D^{\mu} \Phi &= -\frac{\partial V(\Phi)}{\partial \Phi^{\dagger}} = -2\lambda\left(|\Phi|^2 - \frac{v^2}{2}\right)\Phi \, , \\
    D^{\nu} W_{\mu \nu}^a &= J_\mu^a = -i\frac{g}{2}\left[\Phi^{\dagger} \sigma^a \left(D_{\mu} \Phi \right) - \left(D_{\mu} \Phi \right)^{\dagger} \sigma^a \Phi \right] \, , 
\end{align}    
where $D^{\nu} W_{\mu \nu}^a = \partial^{\nu} W_{\mu \nu}^a + g \varepsilon^{abc}W_b^{\nu} W_{\mu \nu c}$ is the gauge-covariant derivative of the field strength tensor, and $J_\mu^a$ represents the Higgs-induced current.

Insertion of our spherically symmetric ansatz into these field equations produces a system of six coupled partial differential equations governing the radial profile functions. The complete set of equations in the presence of both temporal ($G$) and spatial ($f_A$, $f_B$, $f_C$) gauge components, along with the Higgs profiles ($H$, $K$), has been derived in Ref.~\cite{MatchevVerner}. Since our primary focus is sphaleron decay dynamics~\cite{Hellmund:1991ub}, we proceed first with the static sphaleron configuration before examining its real-time evolution using lattice techniques.

For our numerical analysis of sphaleron decay, we adopt the temporal gauge condition $W_0^a = 0$ (equivalently, $G(r,t) = 0$). In this gauge, the dynamical equations for the remaining five profile functions simplify significantly while retaining the essential non-linear couplings that govern the topological transitions. The resulting system is given by:
\begin{widetext}
\begin{equation}
\begin{aligned}
\label{eq:fa1}
\ddot{f}_A - f_A'' 
&+ \frac{1}{r^2} \left(f_A^2 + f_B^2 - 1 \right) f_A 
+ m_W^2 \left[(H^2+ K^2)f_A + K^2 - H^2 \right] + f_A f_C^2 - 2 f_B' f_C - f_B f_C' \; = \; 0 \, ,
\end{aligned}
\end{equation}
\begin{equation}
\begin{aligned}
\label{eq:fb1}
\ddot{f}_B - f_B'' 
&+ \frac{1}{r^2} \left(f_A^2 + f_B^2 - 1 \right) f_B 
+ m_W^2 \left[(H^2+ K^2)f_B - 2HK \right] + f_B f_C^2  + 2 f_A' f_C + f_A f_C' 
\; = \; 0 \, ,
\end{aligned}
\end{equation}
\begin{equation}
\begin{aligned}
\label{eq:fc1}
\ddot{f}_C 
+ \frac{2}{r^2} \left(f_A^2 + f_B^2 \right) f_C 
&+ m_W^2 \left(H^2 + K^2 \right) f_C 
+ 2m_W^2 \left(H'K - HK' \right) 
+ \frac{2}{r^2} \left(f_A' f_B - f_A f_B' \right) \; = \; 0 \, ,
\end{aligned}
\end{equation}
\begin{align}
\label{eq:h1}
   \ddot{H} - \frac{1}{r}(r H)'' +\frac{1}{2r^2} \left(f_A^2 + f_B^2 + 1 \right)H  &- \frac{1}{r^2} \left(H f_A + K f_B \right) + \frac{1}{2}m_H^2 \left(H^2 + K^2 - 1 \right) H ~\nonumber \\
    &- \frac{1}{r} K f_C + \frac{1}{4} H f_C^2 - K'f_C - \frac{1}{2}K f_C'  \; = \; 0 \, ,
\end{align}
\begin{align}
\label{eq:k1}
   \ddot{K} - \frac{1}{r}(r K)'' +\frac{1}{2r^2} \left(f_A^2 + f_B^2 + 1 \right)K  & + \frac{1}{r^2} \left(K f_A - H f_B \right) + \frac{1}{2}m_H^2 \left(H^2 + K^2 - 1 \right) K~\nonumber \\
    &+\frac{1}{r} H f_C + \frac{1}{4} K f_C^2 + H' f_C  + \frac{1}{2}H f_C'\; = \; 0 \, .
\end{align}
\end{widetext}
These equations incorporate the physical values of the $W$-boson mass $m_W = gv/2 = 80.4 \, \rm{GeV}$ and the Higgs boson mass $m_H = \sqrt{2\lambda}v = 125.1 \, \rm{GeV}$~\cite{ParticleDataGroup:2024cfk}. Each term has clear physical significance: the first terms in each equation represent kinetic contributions, the terms involving $1/r^2$ capture the angular momentum barrier, and the remaining terms describe the various gauge-gauge, gauge-Higgs, and Higgs self-interactions. The corresponding equations without gauge fixing, which include the temporal component $G(r,t)$, are presented in Ref.~\cite{MatchevVerner}.
\subsection{Static Sphaleron Configuration}
\label{subsec:staticsolution}
Before investigating the dynamical decay process, we examine the static sphaleron solution that serves as our initial configuration. Our analysis employs a combination of gauge choices optimized for different aspects of the problem: the temporal gauge ($W_0^a = 0$) simplifies the time evolution equations as presented above, while the radial gauge ($x_i W_i^j(x) = 0$) allows for an elegant parametrization of the static sphaleron. These choices are complementary—the former facilitates numerical implementation of dynamics, while the latter makes the topological structure manifest.    

In the radial gauge the sphaleron solution assumes the remarkably simple form~\cite{Manton:1983nd,Klinkhamer:1984di}:
\begin{align}
    f_A(r, t) \; = \; 1 - 2 f(r) \, , \\
    K(r, t) \; = \; h(r) \, , \\
    f_B(r,t) \; = \; f_C(r,t) \; = \; H(r,t) \; = \; G(r, t) \; = \; 0 \, .
\end{align}
This parametrization reduces the system to just two profile functions, $f(r)$ and $h(r)$, which characterize the gauge and Higgs field configurations, respectively. The boundary conditions imposed on these functions are:
\begin{align}
    f(0) &= 0 \,, \quad f(\infty) = 1 \, , \\
    h(0) &= 0 \, , \quad h(\infty) = 1 \, .
\end{align}
These boundary values ensure regularity at the origin while securing the appropriate asymptotic behavior that yields a Chern-Simons number of $\frac{1}{2}$, positioning the sphaleron precisely at the energy barrier between topologically distinct vacua~\cite{Klinkhamer:1984di}.

\begin{figure}[t!]
    \centering
    \includegraphics[width=\linewidth]{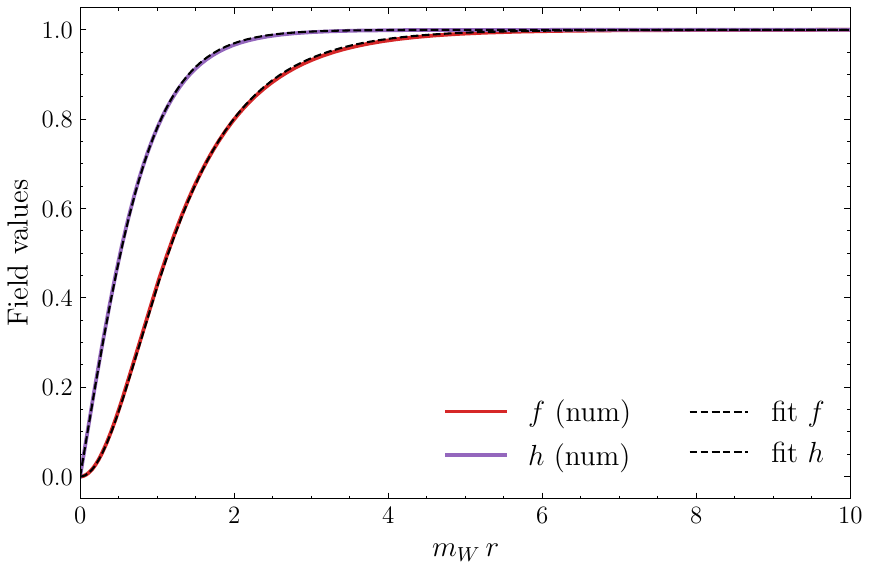}
    \caption{Comparison of the sphaleron radial profile functions $f(r)$ and $h(r)$ obtained through different methods. Solid curves represent results from our numerical constrained energy minimization with fixed Chern-Simons number $N_{CS}=1/2$, while dashed curves show the analytical approximations $f(r) = 1 - \sech(1.15m_Wr)$ and $h(r) = \tanh(1.05 m_W r)$ proposed by Ref.~\cite{Tye:2015tva}. The excellent agreement across all relevant distance scales ($r \sim m_W^{-1}$) validates both approaches.}
    \label{fig:fandhfits}
\end{figure}

For our decay analysis, we employ this parametrization as the initial configuration, subject to perturbations along the unstable direction. The profile functions $f(r)$ and $h(r)$ must generally be determined numerically, but can be accurately approximated by~\cite{Tye:2015tva}:
\begin{equation}
    f(r) = 1 - \sech(1.15m_Wr)\,, \quad h(r) = \tanh(1.05 m_W r) \, .
\end{equation}
These analytical expressions capture the essential features of the sphaleron structure while offering computational advantages for initialization on a discrete lattice. Fig.~\ref{fig:fandhfits} compares our numerically computed profiles with these analytical approximations, confirming excellent agreement across all length scales relevant to the sphaleron dynamics.

To understand the decay pathways, we perform a stability analysis by examining small oscillations around the static solution. Following the approach developed by~\cite{Akiba:1989xu, MatchevVerner}, we parametrize perturbations through a set of fluctuation fields $\eta_i(r,t)$ ($i = B, C, H$):
\begin{align}
\label{eq:gensols}
f_A(r,t) &= 1 - 2f(r) \, ,  \\
f_B(r,t) &= r\,\eta_B(r,t) \, , \\
f_C(r,t) &= \sqrt{2}\,\eta_C(r,t) \, ,  \\
H(r,t)   &= \frac{1}{\sqrt{2}m_W}\,\eta_H(r,t) \, ,  \\
K(r,t)   &= h(r) \, .
\end{align}
This parametrization facilitates dimensional homogeneity and simplifies the resulting linearized equations. When the equations of motion are linearized around the sphaleron solution, they separate into two decoupled sectors: a two-channel sector involving $\eta_A$ and $\eta_K$, and a three-channel sector involving $\eta_B$, $\eta_C$, and $\eta_H$. This decoupling, first noted in Ref.~\cite{Akiba:1989xu}, significantly simplifies the stability analysis. 

Remarkably, the instability resides exclusively in the three-channel ($\eta_B$-$\eta_C$-$\eta_H$) sector. Our analysis, using current SM parameters ($m_H = 125.1$ GeV, $m_W = 80.4$ GeV), confirms the existence of precisely one unstable mode with eigenfrequency $\omega_{-}^2 = -2.7m_W^2$ \cite{Akiba:1989xu}. This negative eigenvalue corresponds to an exponentially growing perturbation that drives the sphaleron decay toward neighboring vacuum configurations.

For numerical analysis, we impose boundary conditions consistent with field regularity at the origin and normalizability at spatial infinity:
\begin{align}
\label{eq:iniconds}
r\eta_i(r)|_{r = 0} &= 0 \, , \quad \text{(regularity at the origin)} \, , \\
\eta_i(r)|_{r = R} &= 0 \, , \quad \text{(normalizability at large $R$)} \, ,
\end{align}
where $R$ represents our numerical cutoff radius, typically $R \gtrsim 50 m_W^{-1}$ in our computations. To solve the resulting eigenvalue problem, we employ a shooting method with adaptive step size, matching solutions at an intermediate point to ensure proper boundary conditions at both $r=0$ and $r=R$. This approach enables precise determination of both the eigenfrequency and the corresponding eigenfunctions. The detailed numerical methodology is described in our companion paper~\cite{MatchevVerner}.

\begin{figure}
    \centering
    \includegraphics[width=1\linewidth]{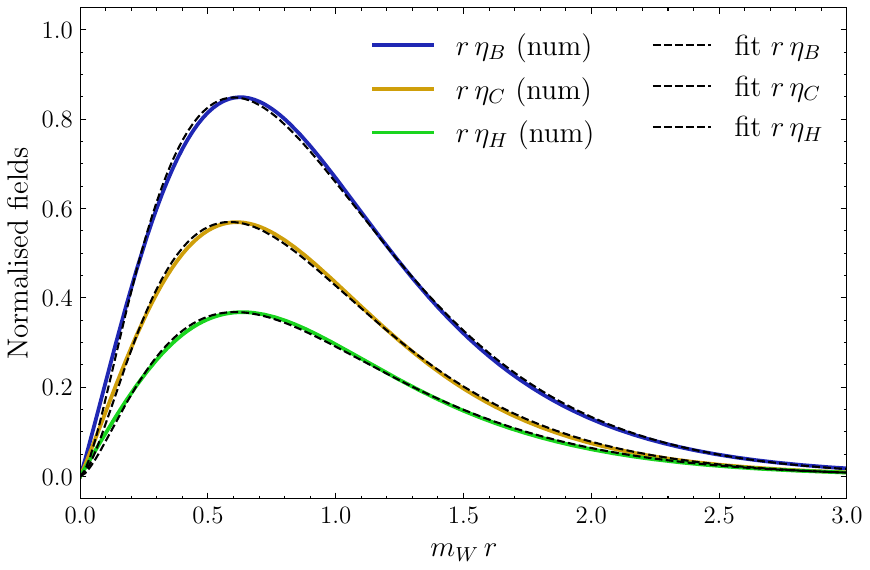}
    \caption{Spatial distribution of the unstable eigenmodes in the three-channel sector, normalized according to Eq.~\eqref{eq:normalization}. The gauge field components $\eta_B(r)$ (blue) and $\eta_C(r)$ (yellow) exhibit more pronounced amplitudes at short distances compared to the Higgs component $\eta_H(r)$ (green), reflecting their different coupling strengths to the background sphaleron configuration. All components display the characteristic exponential localization with effective radius $\sim m_W^{-1}$.}
    \label{fig:normprofiles}
\end{figure}

The unstable mode mediating sphaleron decay exhibits characteristic spatial properties that illuminate the topological transition mechanism. Our analysis yields the following analytical parametrization:
\begin{align}
    \label{eq:etaB_analytic}
    \eta_B(r) &= 10.22\, e^{-2.74\,m_W r} (m_W r)^{0.65} \, , \\
    \label{eq:etaC_analytic}
    \eta_C(r) &= 7.48\, e^{-2.86\,m_W r} (m_W r)^{0.68} \, , \\
    \label{eq:etaH_analytic}
    \eta_H(r) &= 4.00\, e^{-2.62\,m_W r} (m_W r)^{0.59} \, ,
\end{align}
with normalization condition:
\begin{equation}
    \label{eq:normalization}
    \int_0^R dr \, r^2 \left[\eta^2_B(r) + \eta^2_C(r) + \eta^2_H(r) \right] = m_W^{-1} \, .
\end{equation}

As Fig.~\ref{fig:normprofiles} demonstrates, the gauge components dominate at short distances ($r \lesssim m_W^{-1}$), while the Higgs component shows broader spatial distribution. These analytical forms eliminate the need for solving the eigenvalue problem at each lattice site and enable precise control over initial perturbation amplitudes. 

These unstable mode profiles play a crucial role in our lattice investigation of sphaleron decay dynamics. By introducing controlled perturbations along this direction, we can systematically study how the system evolves from the metastable sphaleron configuration toward the topologically distinct vacuum states. The lattice implementation employs a one-dimensional radial grid with spacing $\Delta r \ll m_W^{-1}$ to ensure accurate resolution of the sphaleron structure, while maintaining the essential boundary conditions at $r=0$ and at the lattice boundary. We discuss the details of this implementation in Sec.~\ref{sec:numerical}.

\section{Field Equations Around the Vacuum State}
\label{sec:fieldequationsminimum}
Having established the sphaleron solution and its instability properties, we now examine the field equations around the vacuum state. This analysis serves multiple purposes: it provides a framework for understanding the asymptotic behavior of the sphaleron decay process, establishes the connection to the particle spectrum of the theory, and furnishes the basis for proper boundary conditions in our lattice simulations. The linearized fluctuations around the vacuum constitute the far-field physics of small-amplitude excitations that emerge following the sphaleron decay.

\subsection{Equations of Motion in the Unitary Gauge}
The vacuum configuration in the electroweak theory corresponds to a state where the Higgs field acquires a uniform expectation value and the gauge fields vanish. In terms of our spherically symmetric ansatz, the vacuum is characterized by~\cite{Klinkhamer:1984di}:
\begin{equation}
\begin{aligned}
\label{eq:vacuum_config}
&f_A(r) = 1 \, ,\; f_B(r) = f_C(r) = G(r) = 0 \, ,\\
&H(r) = 1\, ,\; K(r) = 0 \,.
\end{aligned}
\end{equation}
To analyze fluctuations around this vacuum state, we employ the unitary gauge where $K(r,t) = 0$ throughout the evolution. This gauge choice differs from the temporal gauge ($G(r,t) = 0$) used in our sphaleron analysis, but is particularly well-suited for studying small oscillations around the vacuum as it directly corresponds to the physical degrees of freedom. The full equations of motion in this gauge are:
\begin{widetext}
\begin{equation}
\begin{aligned}
\label{eq:fa11}
\ddot{f}_A - f_A'' 
&+ \frac{1}{r^2} \left(f_A^2 + f_B^2 - 1 \right) f_A 
+ m_W^2 H^2(f_A  - 1) + f_A f_C^2 - 2 f_B' f_C - f_B f_C' - f_A G^2 +  2 G \dot{f}_B + f_B \dot{G} = 0 \, ,
\end{aligned}
\end{equation}
\begin{equation}
\begin{aligned}
\label{eq:fb11}
\ddot{f}_B - f_B'' 
&+ \frac{1}{r^2} \left(f_A^2 + f_B^2 - 1 \right) f_B 
+ m_W^2 H^2f_B  + f_B f_C^2  + 2 f_A' f_C + f_A f_C'  - f_B G^2 - 2 G \dot{f}_A  - f_A \dot{G} 
= 0 \, ,
\end{aligned}
\end{equation}
\begin{equation}
\begin{aligned}
\label{eq:fc11}
\ddot{f}_C 
+ \frac{2}{r^2} \left(f_A^2 + f_B^2 \right) f_C 
&+ m_W^2 H^2 f_C 
+ \frac{2}{r^2} \left(f_A' f_B - f_A f_B' \right) + \frac{1}{r}\left(f_B G^2 + 2 G \dot{f}_A +f_A \dot{G} \right) - \dot{G}'  = 0 \, ,
\end{aligned}
\end{equation}
\begin{align}
\label{eq:h11}
    &\ddot{H} - \frac{1}{r}(r H)'' +\frac{1}{2r^2} \left(f_A^2 + f_B^2 + 1 \right)H - \frac{1}{r^2} H f_A + \frac{1}{2}m_H^2 \left(H^2 - 1 \right) H
    + \frac{1}{4} H (f_C^2 - G^2)  \; = \; 0 \, ,
\end{align}
\begin{equation}
\begin{aligned}
\label{eq:g11}
&\frac{1}{r}(r G)''  - \dot{f}_C' - \frac{2}{r} \dot{f}_C 
- \frac{2}{r^2}(f_A^2 + f_B^2) G 
+ \frac{2}{r^2} (f_A \dot{f}_B -  f_B \dot{f}_A ) - m_W^2 H^2 G
 \; = \; 0 \, ,
\end{aligned}
\end{equation}
with the constraint
\begin{align}
\label{eq:k11}
    & \frac{1}{r^2} \left(H f_B \right) -\frac{1}{r} H f_C  - H' f_C  + \dot{H}G - \frac{1}{2}H (f_C' - \dot{G})\; = \; 0 \, .
\end{align}
\end{widetext}

These equations are derived from our original spherically symmetric ansatz by imposing the condition $K(r,t) = 0$. Equation~\eqref{eq:k11} represents the constraint in this gauge, which must be satisfied throughout the evolution to maintain consistency with the gauge condition. 
\subsection{Linearized Fluctuations and Normal Modes}
To analyze small amplitude oscillations around the vacuum state, we expand the fields as:
\begin{align}
\label{eq:fluctuation_expansion}
f_A(r,t) &= 1 + \delta f_A(r,t) \, , \\
f_B(r,t) &= \delta f_B(r,t) \, , \\
f_C(r,t) &= \delta f_C(r,t) \, , \\
H(r,t) &= 1 + \delta H(r,t) \, , \\
G(r,t) &= \delta G(r,t) \, ,
\end{align}
where $\delta f_A$, $\delta f_B$, $\delta f_C$, $\delta H$, and $\delta G$ represent small perturbations. Substituting these expansions into Eqs.~\eqref{eq:fa11}-\eqref{eq:g11} and retaining only terms linear in the fluctuations, we obtain:
\begin{align}
\label{eq:lin_fa}
&\delta \ddot{f}_A - \delta f_A'' + m_W^2 \delta f_A + \frac{2}{r^2}\delta f_A = 0 \, , \\
\label{eq:lin_fb}
&\delta \ddot{f}_B - \delta f_B'' + m_W^2 \delta f_B + \delta f_C' - \delta \dot{G} = 0 \, , \\
\label{eq:lin_fc}
&\delta \ddot{ f}_C + m_W^2 \delta f_C + \frac{2}{r^2} \delta f_C - \frac{2}{r^2} \delta f_B' - \delta \dot{G}' = 0 \, , \\
\label{eq:lin_h}
&\delta \ddot{ H} - \frac{1}{r}(r\delta H)'' + m_H^2 \delta H = 0 \, , \\
\label{eq:lin_g}
&\frac{1}{r} (r\delta G)'' - \delta \dot{f}_C' - \frac{2}{r} \delta \dot{f}_C + \frac{2}{r^2} \delta \dot{f}_B - \frac{2}{r^2} \delta G - m_W^2 \delta G = 0 \, ,
\end{align}
with the linearized constraint
\begin{equation}
\label{eq:lin_constraint}
    \frac{2}{r^2} \delta f_B - \frac{2}{r} \delta f_C + \delta \dot{G} - \delta f_C' = 0 \, .
\end{equation}
This system of equations describes the propagation of small-amplitude waves around the vacuum state. Several important observations can be made. First, Eq.~\eqref{eq:lin_fa} for $\delta f_A$ decouples from the other fluctuations, describing a massive vector mode with mass $m_W$ and an effective ``centrifugal barrier" term $2/r^2$. This corresponds to one component of the massive $W$ bosons in the SM. Second, Eq.~\eqref{eq:lin_h} for $\delta H$ also decouples, describing a massive scalar mode with mass $m_H$, corresponding to the physical Higgs boson. Third, the fields $\delta f_B$, $\delta f_C$, and $\delta G$ are coupled and must be analyzed together. They represent the remaining components of the massive vector bosons.

Using the constraint equation~\eqref{eq:lin_constraint}, we can eliminate $\delta G$ to obtain a coupled system for $\delta f_B$ and $\delta f_C$:
\begin{equation}
\label{eq:fb_reduced}
    \delta \ddot{f}_B - \delta f_B'' + m_W^2 \delta f_B  + \frac{2}{r^2} \delta f_B - \frac{2}{r} \delta f_C = 0 \, ,
\end{equation}
\begin{equation}
\label{eq:fc_reduced}
    \delta \ddot{f}_C + m_W^2 \delta f_C - \frac{4}{r^3}\delta f_B + \frac{4}{r^2} \delta f_C - \frac{2}{r}\delta f_C'-\delta f_C'' = 0 \, .
\end{equation}
This coupled system describes the remaining vector degrees of freedom. To diagonalize it and identify the normal modes, we employ a spherical wave decomposition.

\subsection{Spherical Wave Decomposition}
The linearized fluctuations can be decomposed into normal modes using spherical Bessel functions, which form a natural basis for spherically symmetric wave equations. We express the fluctuations as:
\begin{align}
\label{eq:bessel_fa}
    \frac{\delta f_A(r,t)}{r} &= \int_0^{\infty} \tilde{A}(k, t) j_1(kr) \, dk \, , \\
\label{eq:bessel_fb}
    \frac{\delta f_B(r,t)}{r} &= \int_0^{\infty} \left[\tilde{B}(k, t)j_0(kr)+ \tilde{C}(k,t)j_2(k r) \right] dk \, , \\
\label{eq:bessel_fc}
    \delta f_C(r,t) &= \int_0^{\infty} \left[\tilde{B}(k, t)j_0(kr) - 2\tilde{C}(k,t)j_2(k r) \right] dk \, , \\
\label{eq:bessel_h}
    \delta H(r,t) &= \int_0^{\infty} \tilde{H}(k,t) j_0(kr) dk \, , \\
\label{eq:bessel_g}
    \delta G(r, t) &= \int_0^{\infty} \tilde{G}(k, t)  j_1(k r) dk \, ,
\end{align}
where $j_n(kr)$ represents the spherical Bessel function of order $n$, and $\tilde{A}(k,t)$, $\tilde{B}(k,t)$, $\tilde{C}(k,t)$, $\tilde{H}(k,t)$, and $\tilde{G}(k,t)$ are the corresponding spectral amplitudes that depend on the wavenumber $k$ and time $t$.

Substituting these expressions into Eqs.~\eqref{eq:lin_fa}--\eqref{eq:lin_g} and utilizing the properties of spherical Bessel functions, we find that the spectral amplitudes obey:
\begin{align}
\label{eq:A_spectral}
    \ddot{\tilde{A}}(k,t) + \omega_W^2 \tilde{A}(k,t) &= 0 \, , \\
\label{eq:B_spectral}
    \ddot{\tilde{B}}(k,t) + \omega_W^2 \tilde{B}(k,t) &= 0 \, , \\
\label{eq:C_spectral}
    \ddot{\tilde{C}}(k,t) + \omega_W^2 \tilde{C}(k,t) &= 0 \, , \\
\label{eq:H_spectral}
    \ddot{\tilde{H}}(k,t) + \omega_H^2 \tilde{H}(k,t) &= 0 \, ,
\end{align}
with the dispersion relations:
\begin{equation}
\label{eq:dispersion}
    \omega_W^2 = k^2 + m_W^2 \, , \qquad \omega_H^2 = k^2 + m_H^2 \, .
\end{equation}
The spectral amplitude for $\tilde{G}(k,t)$ is determined from the constraint equation~(\ref{eq:lin_constraint}).

The general solutions for the spectral amplitudes are:
\begin{align}
\label{eq:A_solution}
    &\tilde{A}(k,t) = a(k) \cos(\omega_W t + \varphi_A) \, , \\
\label{eq:B_solution}
    &\tilde{B}(k,t) = \frac{2}{3}b(k) \cos(\omega_W t + \varphi_B) + \frac{1}{3}c(k)\cos(\omega_W t + \varphi_C)  \, , \\
\label{eq:C_solution}
    &\tilde{C}(k,t) = -\frac{1}{3}b(k) \cos(\omega_W t + \varphi_B) + \frac{1}{3}c(k)\cos(\omega_W t + \varphi_C)  \, , \\
\label{eq:H_solution}
    &\tilde{H}(k, t) = h_H(k) \cos(\omega_H t + \varphi_H) \, , \\
\label{eq:G_solution}
    &\tilde{G}(k, t) = \frac{k}{\omega_W} c(k) \sin(\omega_W t + \varphi_C) \, ,
\end{align}
where $a(k)$, $b(k)$, $c(k)$, and $h_H(k)$ represent the spectral amplitudes as functions of wavenumber $k$ only, and $\varphi_A$, $\varphi_B$, $\varphi_C$, and $\varphi_H$ are the corresponding phases. These amplitudes are determined by the initial conditions of the fluctuations.

The inverse transformations, which allow us to compute the spectral amplitudes from the spatial profiles of the fluctuations, are given by:
\begin{align}
\label{eq:inverse_A}
   &\tilde{A}(k,t) = \frac{2k^2}{\pi} \int_0^{\infty} \delta f_A(r,t) r j_1(k r) dr \, , \\
\label{eq:inverse_B}
   &\tilde{B}(k,t) = \frac{2k^2}{\pi} \int_0^{\infty} \frac{1}{3} \left(\frac{2 \delta f_B(r,t)}{r} + \delta f_C(r,t) \right)r^2 j_0(k r) dr \, , \\
\label{eq:inverse_C}
   &\tilde{C}(k,t) = \frac{2k^2}{\pi} \int_0^{\infty} \frac{1}{3}\left(\frac{\delta f_B(r,t)}{r} -\delta f_C(r,t) \right)r^2 j_2(k r) dr \, , \\
\label{eq:inverse_H}
    &\tilde{H}(k,t) = \frac{2k^2}{\pi} \int_0^{\infty} \delta H(r,t) r^2 j_0(k r) dr \, , \\
\label{eq:inverse_G}
     &\tilde{G}(k,t) = \frac{2k^2}{\pi} \int_0^{\infty} \delta G(r,t) r^2 j_1(k r) dr \, .
\end{align}
These transformations are particularly useful for analyzing the asymptotic behavior of the sphaleron decay process. As the sphaleron decays, the energy is redistributed among different normal modes, and the system eventually settles into small oscillations around the vacuum state. By computing the spectral content of these oscillations, we can characterize the efficiency of energy transfer to different wavelengths and particle species.

\subsection{Gauge Transformation Between Different Representations}

In our analysis of the sphaleron (Sec.~\ref{sec:fieldequations}), we employed the temporal gauge ($G(r,t) = 0$), whereas for the vacuum fluctuations, we used the unitary gauge ($K(r,t) = 0$). These different gauge choices are related by the residual $\mathrm{U}(1)$ gauge freedom inherent in our spherically symmetric ansatz. As shown by Ref.~\cite{Ratra:1987dp}, this symmetry acts on the fields as:
\begin{align}
\label{eq:gauge_G}
\begin{pmatrix}
G \\
f_C
\end{pmatrix}
&\rightarrow
\begin{pmatrix}
G \\
f_C
\end{pmatrix}
+
\begin{pmatrix}
\dot{\theta}(r,t) \\
\theta'(r,t)
\end{pmatrix} \, ,  \\
\label{eq:gauge_fA}
\begin{pmatrix}
f_A \\
f_B
\end{pmatrix}
&\rightarrow
\begin{pmatrix}
\cos\theta(r,t) & -\sin\theta(r,t) \\
\sin\theta(r,t) & \cos\theta(r,t)
\end{pmatrix}
\begin{pmatrix}
f_A \\
f_B
\end{pmatrix} \, , \\
\label{eq:gauge_H}
\begin{pmatrix}
H \\
K
\end{pmatrix}
&\rightarrow
\begin{pmatrix}
\cos\left(\frac{\theta(r,t)}{2}\right) & -\sin\left(\frac{\theta(r,t)}{2}\right) \\
\sin\left(\frac{\theta(r,t)}{2}\right) & \cos\left(\frac{\theta(r,t)}{2}\right)
\end{pmatrix}
\begin{pmatrix}
H \\
K
\end{pmatrix} \, , 
\end{align}
where $\theta(r,t)$ is the gauge parameter, with $\theta'(r,t) \equiv \partial \theta(r,t)/\partial r$ and $\dot{\theta}(r,t) \equiv \partial \theta(r,t)/\partial t$.

To transform from the temporal gauge ($G=0$) to the unitary gauge ($K=0$), we need to determine the appropriate gauge function $\theta(r,t)$. For small fluctuations around the vacuum, the required gauge transformation can be determined perturbatively, but for the full non-linear evolution from the sphaleron to the vacuum, numerical integration is generally required.

In our lattice simulations, we implement the following approach. First, we initialize the system in the temporal gauge with the sphaleron configuration, perturbed along its unstable mode. Second, we evolve the system using the equations of motion in the temporal gauge. Finally, to analyze the asymptotic behavior in terms of normal modes,we  transform the final state to the unitary gauge using Eqs.~\eqref{eq:gauge_G}-\eqref{eq:gauge_H} and compute the spectral content using the inverse transformations~\eqref{eq:inverse_A}--\eqref{eq:inverse_G}.

This procedure allows us to systematically study the decay products of the sphaleron and characterize the energy distribution among different particle species (Higgs and gauge bosons) and different wavelengths. In the following section, we will discuss the energy density of the system and analyze its distribution during the sphaleron decay process.

\section{Energy of Decaying Sphaleron}
\label{sec:energyanalysis}

The energy distribution during sphaleron decay provides important insights into the non-perturbative dynamics of topological transitions in the electroweak theory. In this section, we derive the energy functional for our spherically symmetric configuration, analyze its distribution in both coordinate and momentum space, and establish the framework for quantifying particle production during the decay process.

\subsection{General Energy Functional}
\label{subsec:energyfunctional}

The total energy functional for the $SU(2)$-Higgs system can be derived from the Lagrangian density in Eq.~\eqref{eq:SU2lagrangian} through the standard Legendre transformation. For our spherically symmetric ansatz, this energy functional takes the form:
\begin{equation}
    \label{eq:totalenergy}
    E = E_{\rm kin} + E_{\rm pot}\,,
\end{equation}
where $E_{\rm kin}$ represents the kinetic energy contribution and $E_{\rm pot}$ represents the potential (or static) energy contribution.

Working in the unitary gauge where $K(r,t) = 0$, as established in Section~\ref{sec:fieldequationsminimum}, the kinetic energy contribution is given by:
\begin{equation}
\begin{aligned}
    \label{eq:kinen}
    E_{\rm kin} &= \frac{4\pi}{g^2} \int_0^{\infty} dr \Biggl[ \dot{f}_A^2 \left( 1 + \frac{2f_B G}{\dot{f}_A}\right) \\
    &+ \dot{f}_B^2 \left( 1 - \frac{2f_A G}{\dot{f}_B}\right) + \frac{r^2}{2} \dot{f}_C^2 \left(1-2 \frac{G'}{\dot{f}_C} \right) \\
    &+ 2 m_W^2 r^2 \dot{H}^2 \Biggr]\,,
\end{aligned}
\end{equation}
and the potential energy contribution is:
\begin{equation}
\begin{aligned}
    \label{eq:staten}
    E_{\rm pot} &= \frac{4\pi}{g^2} \int_0^{\infty} dr \Biggl[ \left(f_A' + f_C f_B \right)^2 + \left(f_B' - f_C f_A \right)^2 \\
    &+ \frac{\left(f_A^2 + f_B^2 - 1 \right)^2}{2r^2} -\left(f_A^2 + f_B^2\right)G^2 - \frac{1}{2}r^2G'^2 \\
    & + 2 m_W^2 r^2 \bigg\{ H'^2 + \frac{1}{4} f_C^2 H^2 + \frac{1}{2r^2} H^2 (f_A - 1)^2 \\
    &+ \frac{1}{2r^2} H^2 f_B^2 -\frac{1}{4} G^2 H^2 \bigg\} \\
    &+ \frac{(m_W m_H)^2}{2} r^2 \left(H^2 -1 \right)^2 \Biggr]\,.
\end{aligned}
\end{equation}

The derivation of these expressions from the full Lagrangian density is presented in detail in Ref.~\cite{MatchevVerner}. Several important physical features are embedded in these energy functionals. The kinetic terms include non-trivial coupling between time derivatives and the temporal gauge component $G$, reflecting the residual gauge structure. The potential energy includes contributions from gauge field curvature (first two terms), the magnetic energy of non-Abelian fields (third term), gauge-Higgs interactions (terms proportional to $m_W^2$), and the Higgs self-interaction (final term). For the static sphaleron solution presented in Section~\ref{subsec:staticsolution}, where $f_A(r) = 1-2f(r)$, $K(r) = h(r)$, and all other functions vanish, the energy evaluates to $E_{\rm pot} = E_{\rm sph} \simeq 9.0$ TeV, representing the height of the barrier between topologically distinct vacua. During sphaleron decay, this energy is redistributed among various modes, eventually manifesting as propagating gauge and Higgs bosons.

\subsection{Spectral Decomposition of the Energy}
\label{subsec:energyspectral}

To analyze the energy distribution during sphaleron decay, particularly in the asymptotic region where the fields approach small oscillations around the vacuum, we decompose the energy functional in terms of the spectral amplitudes introduced in Section~\ref{sec:fieldequationsminimum}.

We begin by expanding the fields around the vacuum configuration as defined in Eq.~\eqref{eq:fluctuation_expansion} and retaining terms up to quadratic order in the fluctuations. The resulting expression for the kinetic energy is:
\begin{equation}
\begin{aligned}
   \label{eq:kin_quadratic}
   E_{\rm kin}^{(2)} &= \frac{4\pi}{g^2} \int_0^{\infty} dr \Biggl[ \delta \dot{f}_A^2 + \delta \dot{f}_B^2 + \frac{r^2}{2}\delta \dot{f}_C^2 \\
   &- 2\delta G\delta \dot{f}_B - r^2\delta G'\delta \dot{f}_C + 2m_W^2 r^2 \delta \dot{H}^2 \Biggr]\,.
\end{aligned}
\end{equation}

Similarly, the quadratic approximation to the potential energy is:
\begin{equation}
\begin{aligned}
   \label{eq:pot_quadratic}
   &E_{\rm pot}^{(2)} = \frac{4\pi}{g^2} \int_0^{\infty} dr \Biggl[ (\delta f_A')^2 + (\delta f_B')^2 + \frac{2}{r^2}(\delta f_A)^2 \\
   &+ m_W^2\delta f_A^2 + m_W^2\delta f_B^2 + \frac{r^2}{2}m_W^2 \delta f_C^2 - 2 \delta f_C \delta f_B' - \frac{r^2}{2}\delta G'^2 \\
   &- \delta G^2 - \frac{r^2}{2} m_W^2\delta G^2 + 2m_W^2 r^2 (\delta H')^2 + 2m_W^2 m_H^2 r^2 \delta H^2 \Biggr]\,.
\end{aligned}
\end{equation}

Using the spherical wave decomposition introduced in Eqs.~\eqref{eq:bessel_fa}--\eqref{eq:bessel_g}, and applying the orthogonality properties of spherical Bessel functions, we can express the energy in terms of the spectral amplitudes. For the kinetic energy, we obtain:
\begin{equation}
\begin{aligned}
   \label{eq:kin_spectral}
   &E_{\rm kin}^{(2)}(t) = \frac{2\pi^2}{g^2} \int_0^{\infty} \frac{dk}{k^2} \bigg[\dot{\tilde{A}}^2(k,t) + \frac{3}{2}\dot{\tilde{B}}^2(k,t) \\
   &+ 3\dot{\tilde{C}}^2(k,t) + 2m_W^2\dot{\tilde{H}}^2(k,t) \\
   &- k \big(\dot{\tilde{B}}(k,t) \tilde{G}(k,t) + 2 \dot{\tilde{C}}(k,t) \tilde{G}(k,t)\big) \bigg]\,,
\end{aligned}    
\end{equation}
and for the potential energy:
\begin{equation}
\begin{aligned}
   \label{eq:pot_spectral}
   &E_{\rm pot}^{(2)}(t) = \frac{2\pi^2}{g^2} \int_0^{\infty} \frac{dk}{k^2} \bigg[(m_W^2 + k^2) \tilde{A}^2(k,t)  \\
   &+\left(\frac{3m_W^2}{2} + k^2 \right)\tilde{B}^2(k,t) + \left(3 m_W^2 + k^2 \right) \tilde{C}^2(k,t) \\
   &-\left(\frac{m_W^2 + k^2}{2} \right) \tilde{G}^2(k,t) - 2k^2 \tilde{B}(k,t) \tilde{C}(k,t) \\
   &+ 2k^2 m_W^2 \tilde{H}^2(k,t) + 2 m_W^2 m_H^2 \tilde{H}^2(k,t)  \bigg]\,.
\end{aligned}    
\end{equation}

These expressions provide the spectral decomposition of the energy in terms of the mode amplitudes in $k$-space. While they appear complex, they can be significantly simplified by using the dynamical equations for the spectral amplitudes derived in Section~\ref{sec:fieldequationsminimum}.

\subsection{Energy in Terms of Normal Modes}
\label{subsec:energymodes}

The expressions for the energy can be further simplified by using the relations between different spectral amplitudes established by the constraint equation~\eqref{eq:lin_constraint} and the solutions given by Eqs.~\eqref{eq:A_solution}--\eqref{eq:G_solution}. 

Substituting these solutions into Eqs.~\eqref{eq:kin_spectral} and \eqref{eq:pot_spectral}, and combining the kinetic and potential energy contributions, we obtain a remarkably simple expression for the total energy:
\begin{equation}
\begin{aligned}
   \label{eq:total_energy_modes}
   E(t) &= \frac{2\pi^2}{g^2} \int_0^{\infty} \frac{dk}{k^2} \bigg[ \omega_W^2 a^2(k) + \omega_W^2 b^2(k) \\
   &+ \frac{1}{2}m_W^2 c^2(k) + 2 \omega_H^2 m_W^2 h_H^2(k) \bigg]\,.
\end{aligned}    
\end{equation}

This elegant form reveals the physical interpretation of the spectral amplitudes. The $a(k)$ amplitude represents a transverse gauge field mode with dispersion relation $\omega_W^2 = k^2 + m_W^2$, corresponding to one polarization state of the massive $W$ boson. The $b(k)$ amplitude represents another transverse gauge field mode with the same dispersion relation, providing the second polarization state of the $W$ boson. The $c(k)$ amplitude corresponds to the longitudinal mode of the $W$ boson, which arises from the ``eaten" Goldstone boson in the Higgs mechanism. The $h_H(k)$ amplitude represents the physical Higgs boson mode with dispersion relation $\omega_H^2 = k^2 + m_H^2$. The factor of $1/2$ in front of the $c(k)$ term is a consequence of the particular normalization chosen for the longitudinal mode in our decomposition. The overall prefactor $2\pi^2/g^2$ accounts for the coupling constant dependence and the spherical symmetry of our configuration.

\subsection{Particle Number and Energy Distribution}
\label{subsec:particlenumber}

One of the primary objectives of our analysis is to determine the number and energy distribution of particles produced during sphaleron decay. The spectral energy density derived above provides the natural framework for this analysis.

For each normal mode, we can define a spectral energy density:
\begin{equation}
\begin{aligned}
   \label{eq:spectral_densities}
   \frac{dE_W^T(k)}{dk} &= \frac{2\pi^2}{g^2} \frac{\omega_W^2}{k^2} \big(a^2(k) + b^2(k)\big)\,, \\
   \frac{dE_W^L(k)}{dk} &= \frac{2\pi^2}{g^2} \frac{m_W^2}{2k^2} c^2(k)\,, \\
   \frac{dE_H(k)}{dk} &= \frac{2\pi^2}{g^2} \frac{2\omega_H^2 m_W^2}{k^2} h_H^2(k)\,,
\end{aligned}    
\end{equation}
where $E_W^T$ represents the energy in transverse $W$ boson modes, $E_W^L$ represents the energy in longitudinal $W$ boson modes, and $E_H$ represents the energy in Higgs boson modes.

The corresponding particle number densities can be obtained by dividing the energy densities by the energy per particle ($\hbar\omega$):
\begin{equation}
\begin{aligned}
   \label{eq:number_densities}
   \frac{dN_W^T(k)}{dk} &= \frac{2\pi^2}{g^2} \frac{\omega_W}{k^2} \big(a^2(k) + b^2(k)\big)\,, \\
   \frac{dN_W^L(k)}{dk} &= \frac{2\pi^2}{g^2} \frac{m_W^2}{2\omega_W k^2} c^2(k)\,, \\
   \frac{dN_H(k)}{dk} &= \frac{2\pi^2}{g^2} \frac{2\omega_H m_W^2}{k^2} h_H^2(k)\,.
\end{aligned}    
\end{equation}

These spectral densities enable us to quantify the particle production process during sphaleron decay. By numerically evolving the full non-linear field equations from the perturbed sphaleron initial state and analyzing the spectral content of the final state, we can determine the total number of particles produced in each species, the momentum distribution of these particles, and the energy conversion efficiency from the initial sphaleron energy to various particle species.

In the following section, we present our numerical implementation and the results of our lattice simulations of sphaleron decay dynamics.

\section{Numerical Analysis of Sphaleron Decay}
\label{sec:numerical}
Having established the theoretical framework for sphaleron decay and its energy distribution, we now present a systematic numerical investigation of the decay dynamics. This section describes our lattice implementation, numerical procedures, and the results of simulations under various initial conditions that trigger the decay process.

To accurately capture the sphaleron decay dynamics, we implement a high-resolution lattice discretization of the coupled partial differential equations derived in Section~\ref{sec:fieldequations}. Specifically, we discretize Eqs.~\eqref{eq:fa1}--\eqref{eq:k1} on a one-dimensional radial grid with the following specifications:

\begin{itemize}
    \item Spatial lattice: $N_r = 5000$ points spanning from $r = 0$ to $r = 50 m_W^{-1}$, with uniform spacing $\Delta r = 0.001 m_W^{-1}$.
    \item Temporal lattice: Evolution from $t = 0$ to $t = 50 m_W^{-1}$, with time step $\Delta t = 0.00001 m_W^{-1}$.
    \item Physical parameters: $m_W = 80.4 \, \rm{GeV}$ and $m_H = 125.1 \, \rm{GeV}$, corresponding to the measured values in the Standard Model~\cite{ParticleDataGroup:2024cfk}.
\end{itemize}

For the spatial discretization, we employ second-order accurate central difference approximations for the radial derivatives, with special attention to the boundary conditions at $r = 0$ and $r = R = 50 m_W^{-1}$. At the origin, we impose regularity conditions and at the outer boundary $R$, we impose the normalizability at large $R$ condition, given by Eq.~(\ref{eq:iniconds}). The time evolution is performed using a fourth-order Runge-Kutta method, which provides excellent accuracy and stability. The constraints are monitored throughout the evolution, with deviations maintained below $10^{-10}$ relative to the total energy. To further validate our numerical approach, we verify energy conservation finding that the total energy is preserved to better than $0.1\%$ over the entire simulation period.\footnote{Our complete implementation is publicly available on GitHub at \url{https://github.com/sarunasverner/sphaleron-decay}.}

The sphaleron, being a saddle point in configuration space, can decay along its unstable direction through various mechanisms. To systematically explore the dynamics and resulting particle production, we investigate four distinct scenarios characterized by different initial conditions:

\begin{enumerate}
    \item \textbf{Case I: Small displacement with no initial kinetic energy.} The system is displaced slightly along the unstable direction, allowing for a controlled study of the decay process when triggered by potential energy.
    
    \item \textbf{Case II: Large displacement with no initial kinetic energy.} A more substantial displacement is imposed, examining how the decay characteristics scale with the magnitude of the perturbation.
    
    \item \textbf{Case III: No displacement with initial kinetic energy.} The sphaleron configuration is given an initial velocity along the unstable direction, modeling scenarios where the decay is triggered by dynamical processes rather than configuration perturbations.
    
    \item \textbf{Case IV: Small displacement with initial kinetic energy.} A combination of both displacement and initial momentum, representing a more general decay scenario.
\end{enumerate}

These scenarios are physically motivated by different processes that might trigger sphaleron transitions in early universe cosmology or high-energy collisions. The first two cases model thermal fluctuations of varying magnitudes, while the third case corresponds to dynamical excitation through interactions with other fields. The fourth case represents the most general scenario where both effects are present.

\subsection{Case I: Small Displacement with No Kinetic Energy}
\label{subsec:case1}
\begin{figure*}[ht!]
    \centering
    \begin{minipage}{0.48\linewidth}
        \centering
        \includegraphics[width=\linewidth]{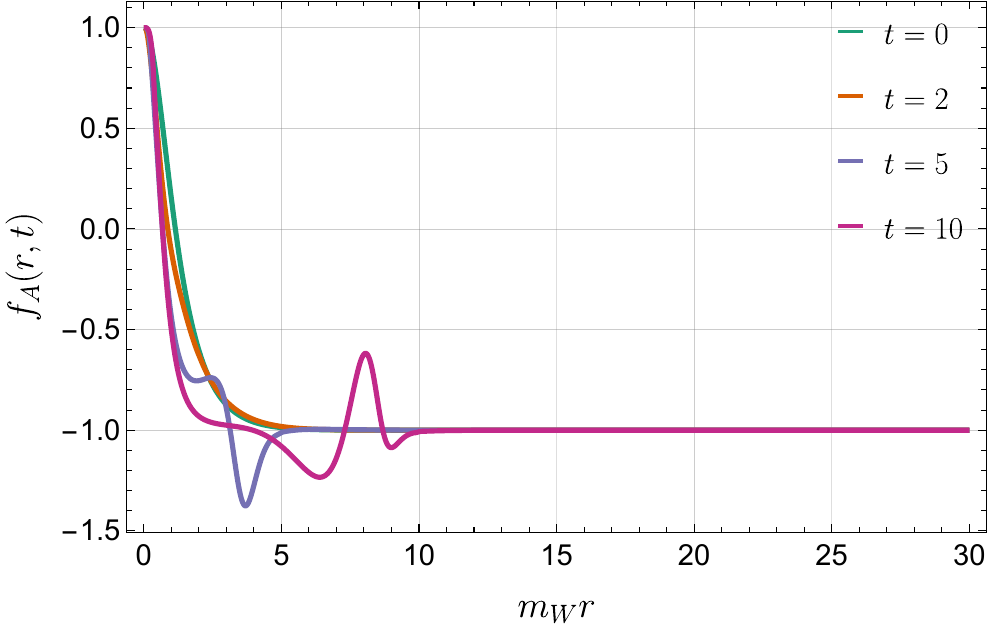}
    \end{minipage}
    \hfill
    \begin{minipage}{0.48\linewidth}
        \centering
        \includegraphics[width=\linewidth]{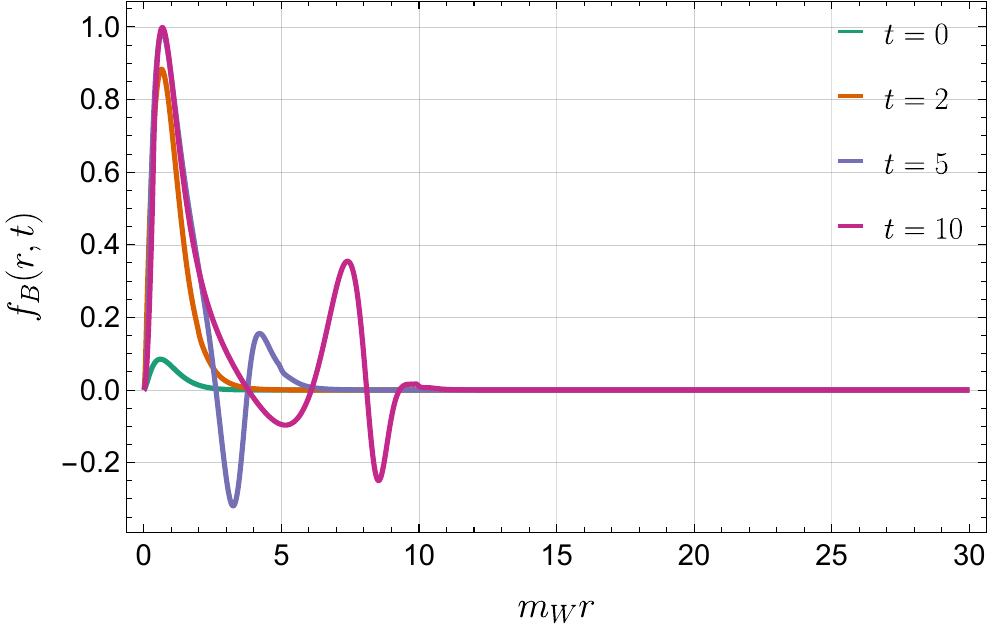}
    \end{minipage}
    
    \vspace{1em}
    
    \begin{minipage}{0.48\linewidth}
        \centering
        \includegraphics[width=\linewidth]{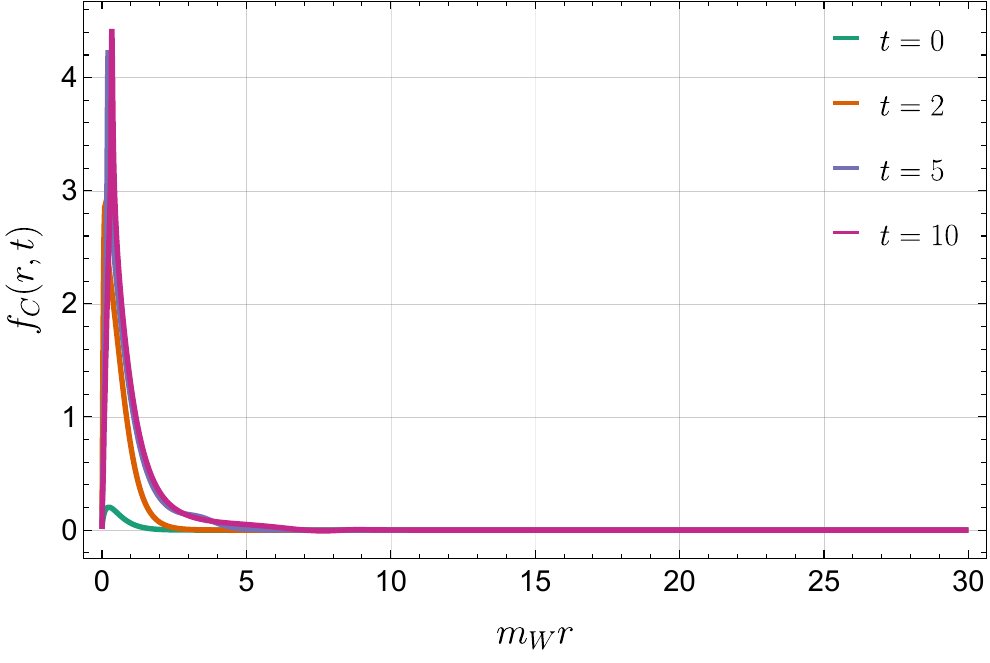}
    \end{minipage}
    \hfill
    \begin{minipage}{0.48\linewidth}
        \centering
        \includegraphics[width=\linewidth]{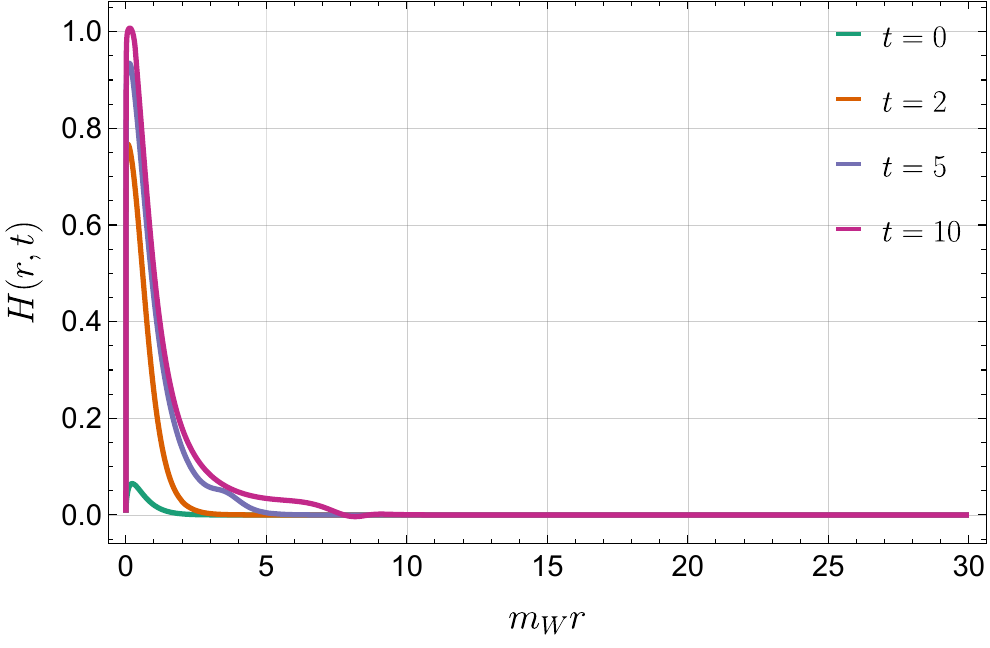}
    \end{minipage}
    
    \vspace{1em}
    
    \begin{minipage}{0.48\linewidth}
        \centering
        \includegraphics[width=\linewidth]{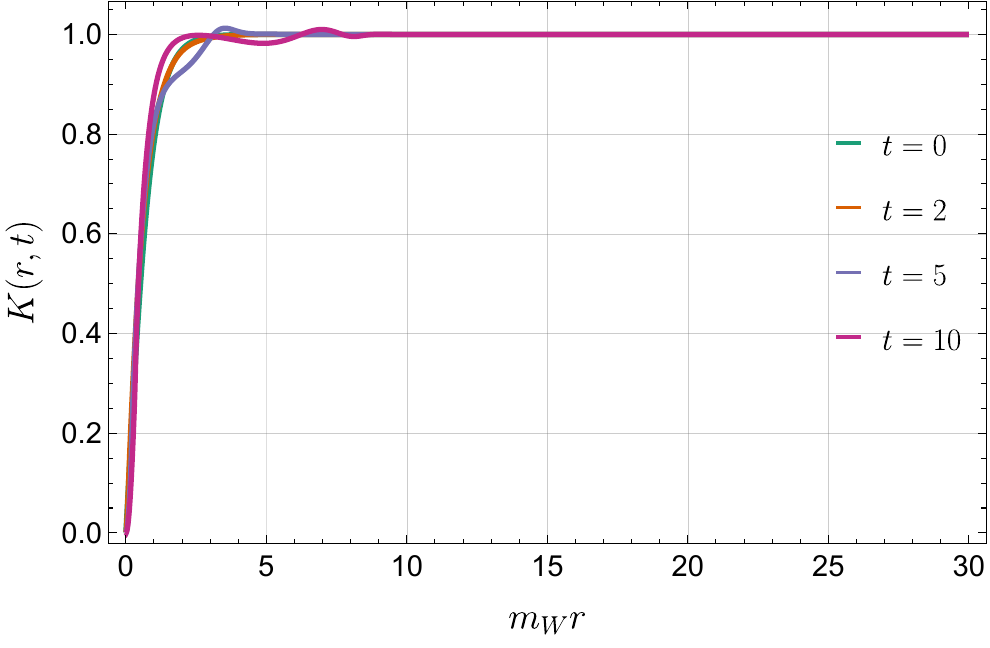}
    \end{minipage}
    \hfill
    \begin{minipage}{0.48\linewidth}
        \centering
        \includegraphics[width=\linewidth]{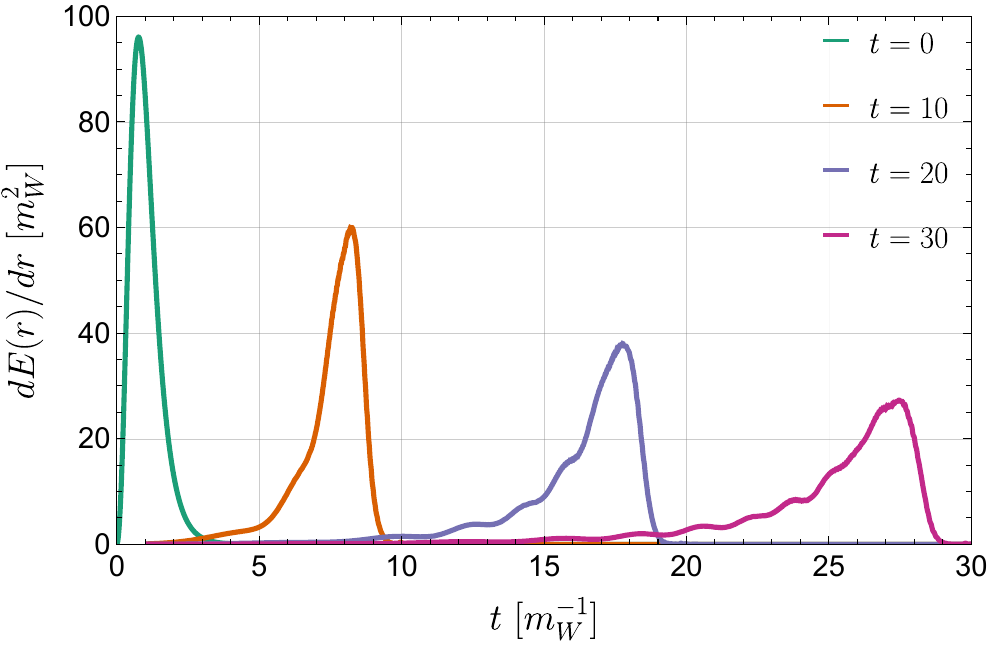}
    \end{minipage}

    \caption{Evolution of field profiles during sphaleron decay for Case I (small displacement, no initial kinetic energy). The top and middle rows show the profiles of $f_A$, $f_B$, $f_C$, and $H$ as functions of $m_W r$ at times $t = 0$, $2 m_W^{-1}$, $5 m_W^{-1}$, and $10 m_W^{-1}$. The bottom left panel shows the profile of $K$. The bottom right panel displays the radial energy density distribution $dE/dr$ in units of $m_W^2$ at times $t = 0$, $10 m_W^{-1}$, $20 m_W^{-1}$, and $30 m_W^{-1}$, illustrating how the initially localized energy propagates outward as the sphaleron decays.}
    \label{fig:sphaleron_funcs_1}
\end{figure*}

\begin{figure*}[t!]
    \centering
    \begin{minipage}{0.47\linewidth}
        \centering
        \includegraphics[width=\linewidth]{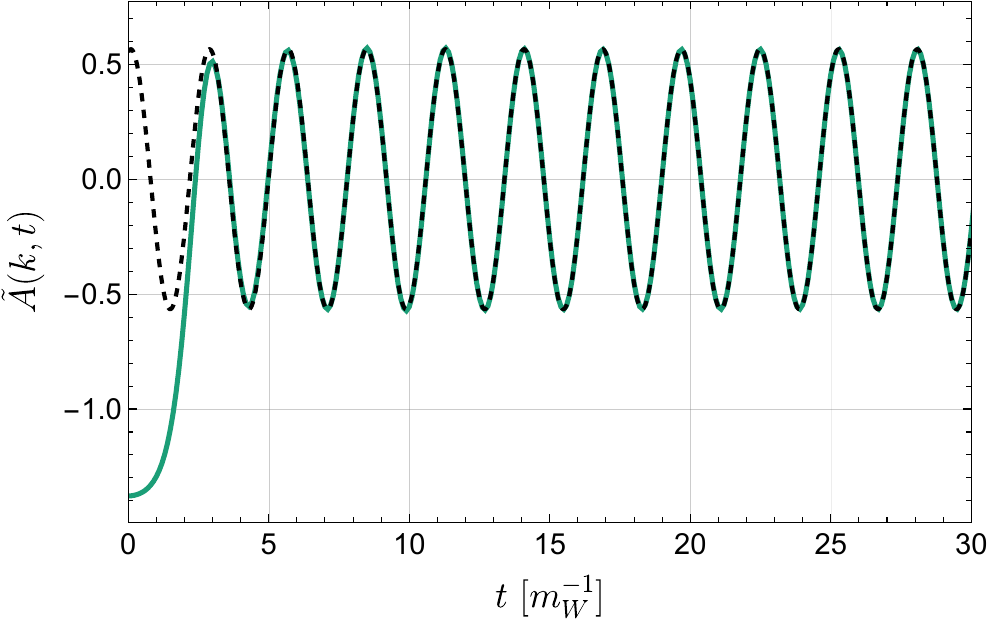}
    \end{minipage}
    \hfill
    \begin{minipage}{0.47\linewidth}
        \centering
        \includegraphics[width=\linewidth]{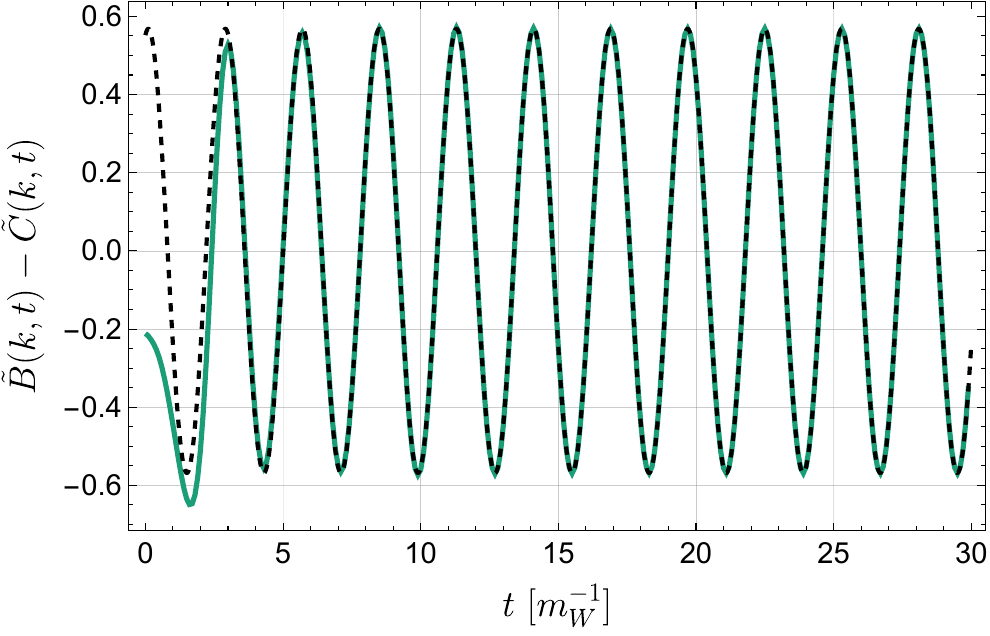}
    \end{minipage}
    
    \vspace{1em}
    
    \begin{minipage}{0.47\linewidth}
        \centering
        \includegraphics[width=\linewidth]{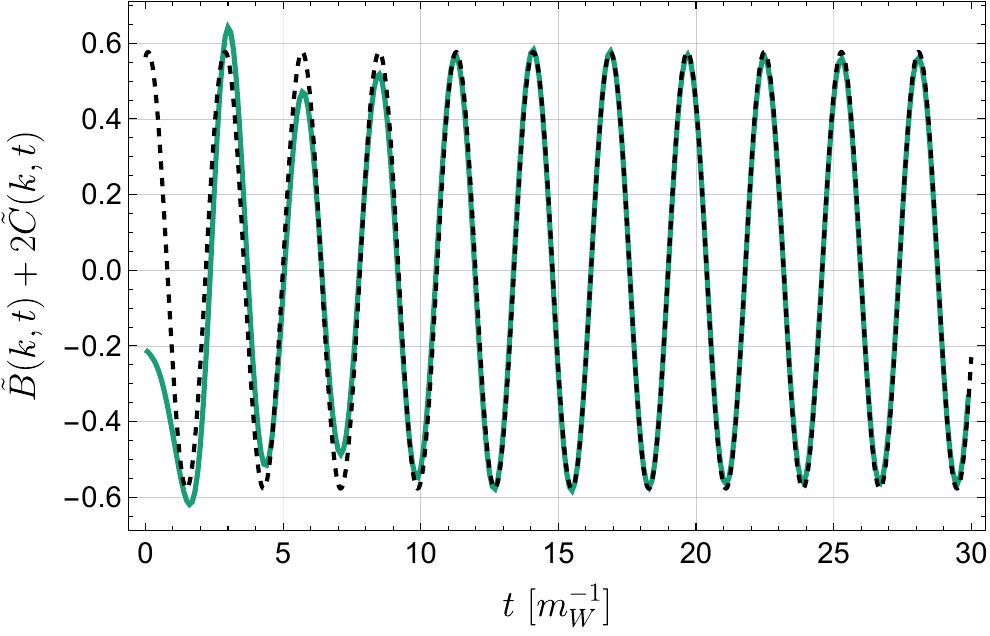}
    \end{minipage}
    \hfill
    \begin{minipage}{0.47\linewidth}
        \centering
        \includegraphics[width=\linewidth]{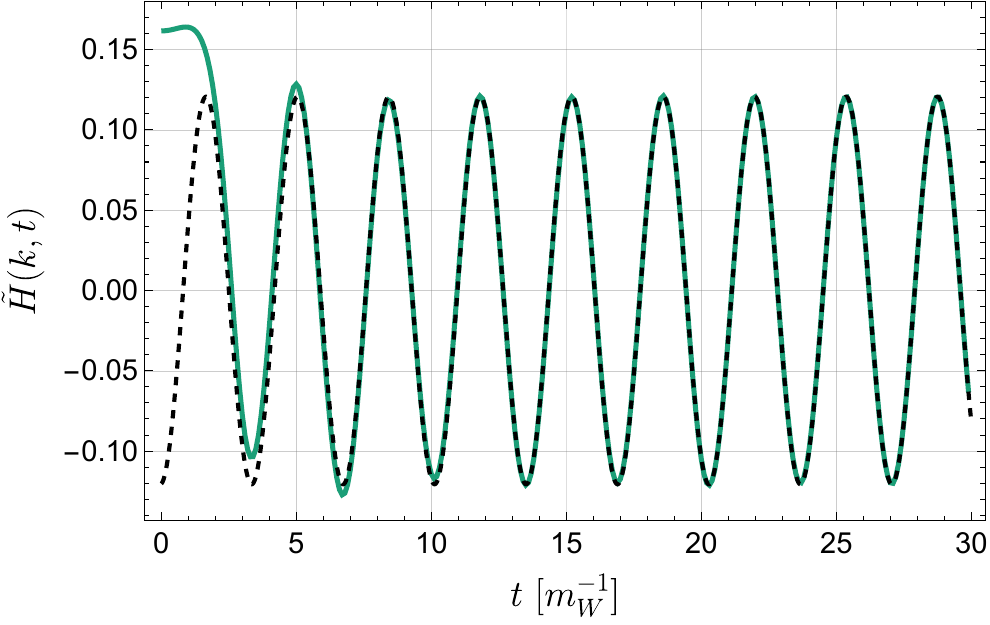}
    \end{minipage}
    \caption{Time evolution of spectral amplitudes at fixed momentum $k = 2m_W$ for Case I. The plots show the Fourier modes $\tilde{A}(k,t)$ (top left), $\tilde{B}(k,t) - \tilde{C}(k,t)$ (top right), $\tilde{B}(k,t) + 2\tilde{C}(k,t)$ (bottom left), and $\tilde{H}(k,t)$ (bottom right). After the initial transient phase ($t \gtrsim 10 m_W^{-1}$), the modes exhibit the expected free-field oscillatory behavior with frequencies $\omega_W = \sqrt{k^2 + m_W^2}$ for gauge fields and $\omega_H = \sqrt{k^2 + m_H^2}$ for the Higgs field. The dashed black curves show fits to these oscillations, from which we extract the amplitudes $a(k)$, $b(k)$, $c(k)$, and $h_H(k)$ used to compute particle multiplicities.}
    \label{fig:sphaleron_fourier_1}
\end{figure*}

We begin our analysis with a small displacement along the unstable direction identified in Section~\ref{sec:fieldequations}. Specifically, we set the initial conditions as:
\begin{align}
    \label{eq:case1_init1}
    f_B(r, 0) &= 0.1 \times r\, \eta_B(r) \, , \\
    \label{eq:case1_init2}
    f_C(r, 0) &= 0.1 \times \sqrt{2} \, \eta_C(r) \, , \\
    \label{eq:case1_init3}
    H(r, 0) &= 0.1 \times \frac{1}{\sqrt{2} m_W}  \, \eta_H(r) \, , 
\end{align}
where $\eta_B(r)$, $\eta_C(r)$, and $\eta_H(r)$ are the components of the unstable eigenvector with normalization given by Eq.~(\ref{eq:normalization}), leading to 
\begin{equation}
    \label{eq:case1_norm}
    \int_0^R dr \, r^2 \times 0.01 \left[\eta^2_B(r) + \eta^2_C(r) + \eta^2_H(r) \right] = 0.01 m_W^{-1} \, ,
\end{equation}
while $f_A(r, 0) = 1 - 2f(r)$ and $K(r, 0) = 0$ according to our gauge choice. We set all initial velocities to zero:
\begin{align}
    \label{eq:case1_vel}
    \dot{f}_A(r,0) = \dot{f}_B(r,0) = \dot{f}_C(r,0) = \dot{H}(r,0) = \dot{K}(r,0) = 0 \, .
\end{align}

The factor of 0.1 in Eqs.~\eqref{eq:case1_init1}--\eqref{eq:case1_init3} is chosen to ensure that the initial perturbation is small enough to allow for a controlled and smooth decay process. The corresponding initial energy is approximately $0.1\%$ below the sphaleron energy $E_{\mathrm{sph}} \simeq 9.1$ TeV.

Figure~\ref{fig:sphaleron_funcs_1} shows the evolution of the field profiles at several representative times during the decay process. In the early stages ($t \lesssim 2 m_W^{-1}$), the fields evolve primarily along the unstable direction, with the displacement amplifying exponentially as expected from linear stability analysis. By $t \simeq 5 m_W^{-1}$, the system has evolved significantly away from the sphaleron configuration.

The bottom-right panel of Fig.~\ref{fig:sphaleron_funcs_1} depicts the radial energy density distribution at various times, revealing how the initially localized energy of the sphaleron gradually propagates outward as the system decays. By $t \simeq 30 m_W^{-1}$, most of the energy has been converted into outgoing waves of gauge and Higgs bosons.

To analyze the spectral content of these outgoing waves, we track the time evolution of the spectral amplitudes introduced in Section~\ref{sec:fieldequationsminimum}. Figure~\ref{fig:sphaleron_fourier_1} shows the time evolution of the Fourier modes $\tilde{A}(k,t)$, $\tilde{B}(k, t) - \tilde{C}(k, t)$, $\tilde{B}(k, t) + 2\tilde{C}(k, t)$, and $\tilde{H}(k, t)$ for a representative momentum $k = 2 m_W$, which lies near the peak of the spectrum.

Initially, the spectral amplitudes evolve in a complex, non-linear manner as the sphaleron decays. However, after sufficient time ($t \gtrsim 5 m_W^{-1}$), the system enters a free-field regime where the spectral amplitudes oscillate with their characteristic frequencies $\omega_W = \sqrt{k^2 + m_W^2}$ for gauge bosons and $\omega_H = \sqrt{k^2 + m_H^2}$ for Higgs bosons. By fitting these late-time oscillations to the forms given in Eqs.~\eqref{eq:A_solution}--\eqref{eq:H_solution}, we extract the amplitudes $a(k)$, $b(k)$, $c(k)$, and $h_H(k)$.

\begin{figure}[h!]
    \centering
    \includegraphics[width=1\linewidth]{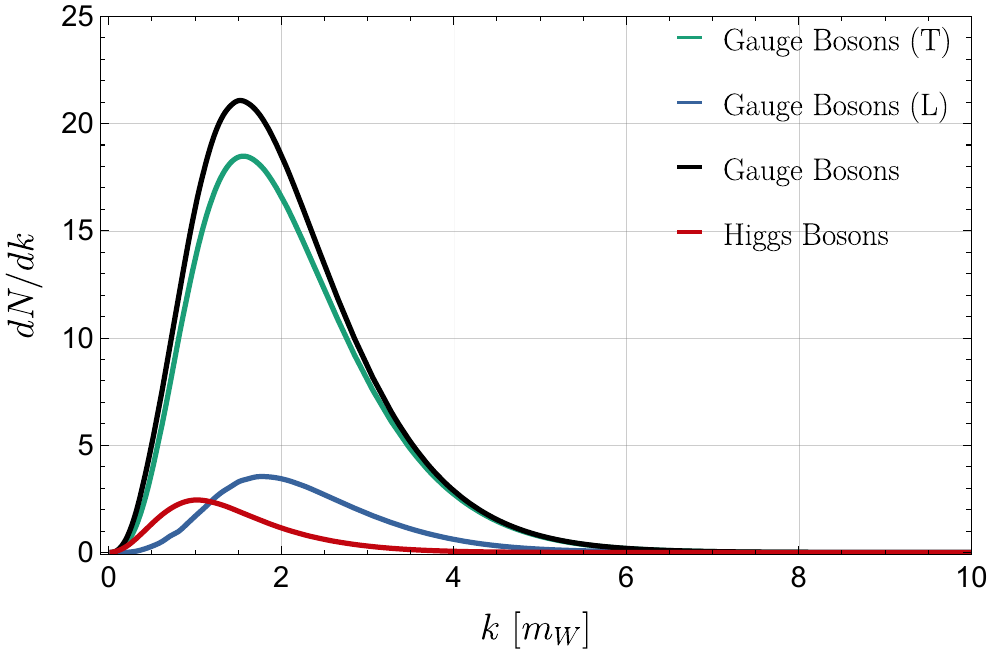}
    \caption{Particle multiplicity distributions $dN/dk$ as functions of momentum $k$ (in units of $m_W$) for Case I. The curves show the distributions for longitudinal gauge bosons (blue), transverse gauge bosons (green), and Higgs bosons (red). The transverse mode dominates the spectrum, particularly at lower momenta, reflecting the direct connection between this mode and the topological structure of the sphaleron. The distributions peak around $k \simeq 1.5 m_W$ for gauge bosons and $k \simeq 1 m_W$ for Higgs bosons, consistent with the characteristic size of the sphaleron ($\sim m_W^{-1}$).}
    \label{fig:multiplicity1}
\end{figure}

These amplitudes allow us to compute the particle number distributions according to Eq.~\eqref{eq:number_densities}. Figure~\ref{fig:multiplicity1} presents the resulting multiplicity distributions for the different particle species produced during the sphaleron decay.

The multiplicity distributions reveal several important features of the particle production process. The spectrum is dominated by transverse gauge bosons, with a total multiplicity of 41.6 compared to 7.8 for longitudinally polarized gauge bosons (for a total of 49.4 gauge bosons) and 4.0 for Higgs bosons. The distributions peak at momenta $k \simeq 1.5 m_W$ for gauge bosons and $k \simeq 1 m_W$ for Higgs bosons, corresponding to wavelengths comparable to the characteristic size of the sphaleron ($\sim m_W^{-1}$). The spectral shapes are approximately exponential at high momenta, reminiscent of a thermal-like distribution, despite the fact that the decay process is entirely classical and deterministic.

The dominance of transverse gauge bosons in the spectrum reflects the nature of the gauge field dynamics during the sphaleron decay. While the longitudinal polarization states (originating from the "eaten" Goldstone bosons) are connected to the topological structure, the sphaleron decay predominantly excites the transverse degrees of freedom that carry the majority of the field energy during the transition process.

The total energy distribution among different particle species reveals that Higgs bosons account for only 7.3\% of the total energy, with the remainder divided between transverse (approximately 78\%) and longitudinal (approximately 14.7\%) gauge bosons. This asymmetric energy distribution reflects the specific nature of the unstable mode that triggers the decay, which couples more strongly to the transverse gauge field components than to the scalar field sector.
\subsection{Case II: Large Displacement with No Kinetic Energy}
\label{subsec:case2}
\begin{figure*}[t!]
    \centering
    \begin{minipage}{0.48\linewidth}
        \centering
        \includegraphics[width=\linewidth]{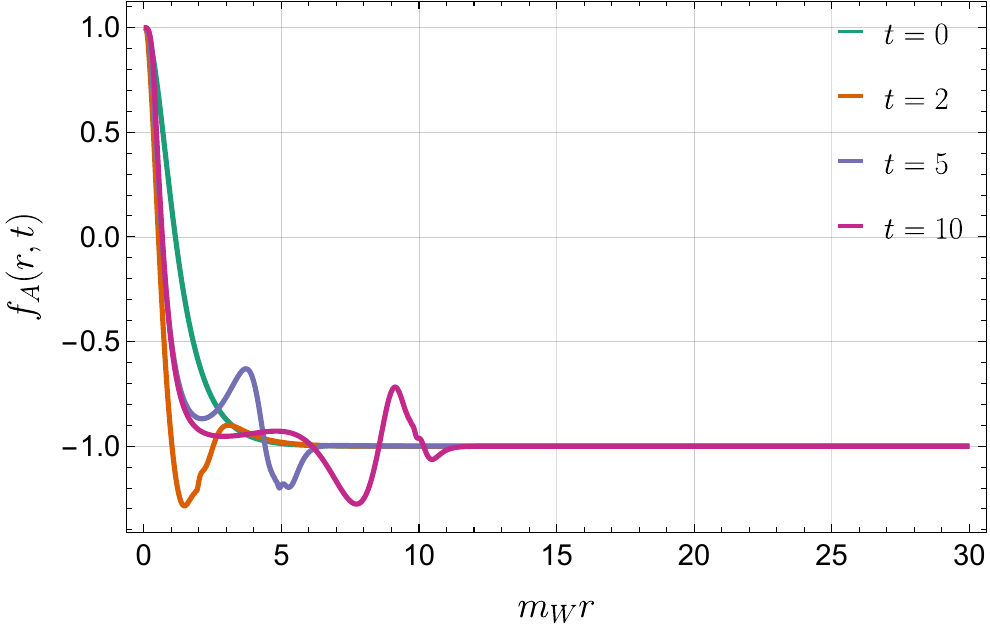}
    \end{minipage}
    \hfill
    \begin{minipage}{0.48\linewidth}
        \centering
        \includegraphics[width=\linewidth]{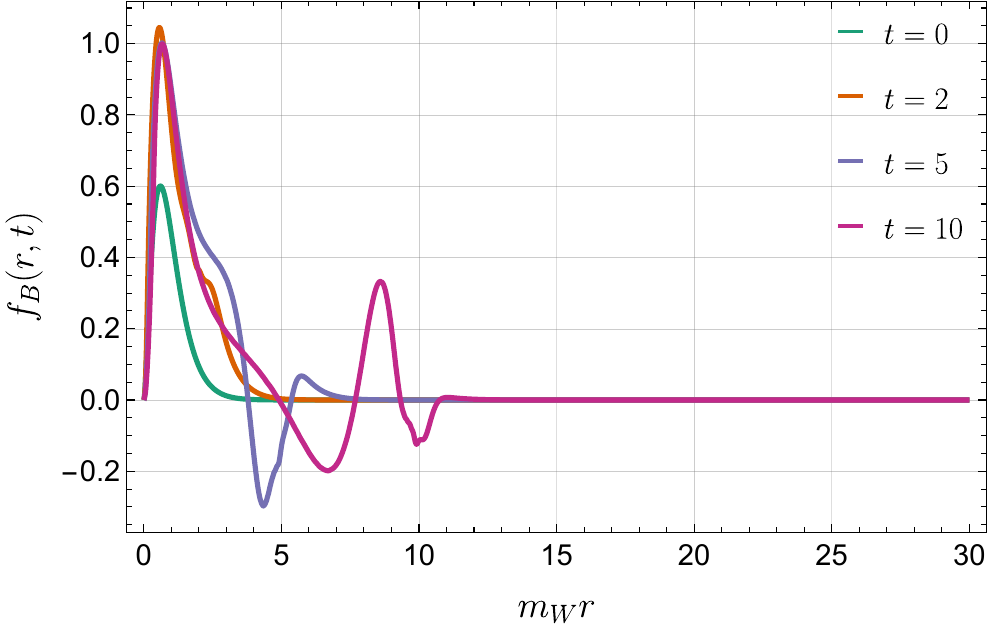}
    \end{minipage}
    
    \vspace{1em}
    
    \begin{minipage}{0.48\linewidth}
        \centering
        \includegraphics[width=\linewidth]{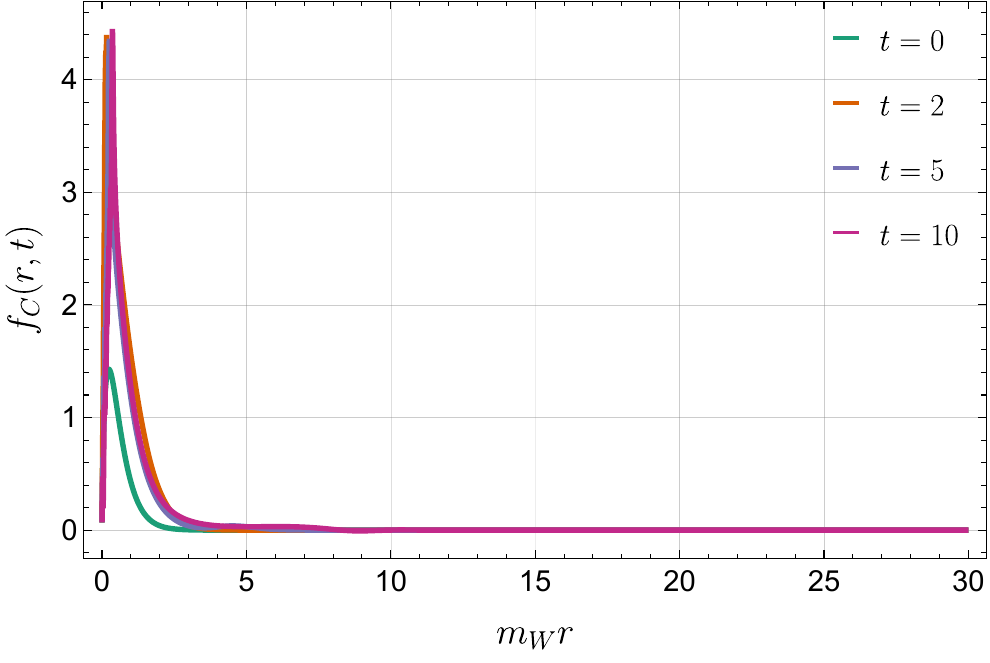}
    \end{minipage}
    \hfill
    \begin{minipage}{0.48\linewidth}
        \centering
        \includegraphics[width=\linewidth]{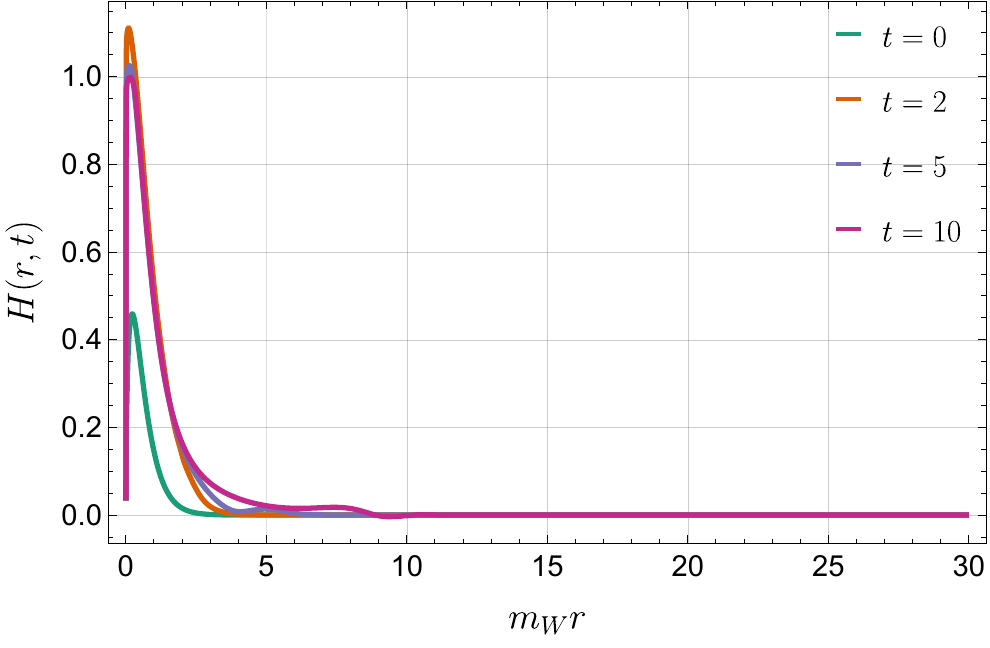}
    \end{minipage}
    
    \vspace{1em}
    
    \begin{minipage}{0.48\linewidth}
        \centering
        \includegraphics[width=\linewidth]{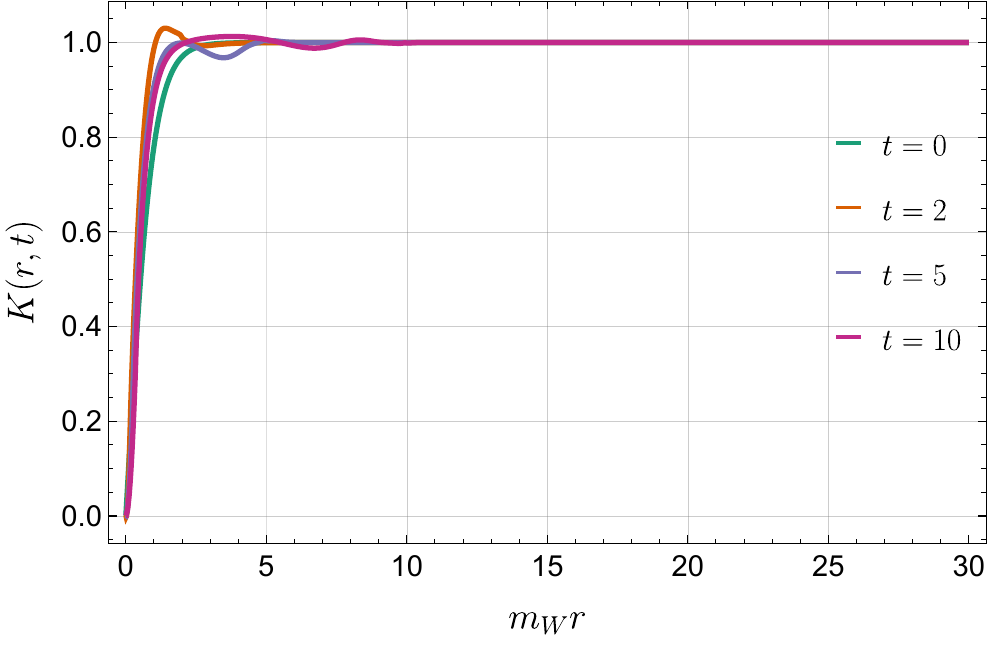}
    \end{minipage}
    \hfill
    \begin{minipage}{0.48\linewidth}
        \centering
        \includegraphics[width=\linewidth]{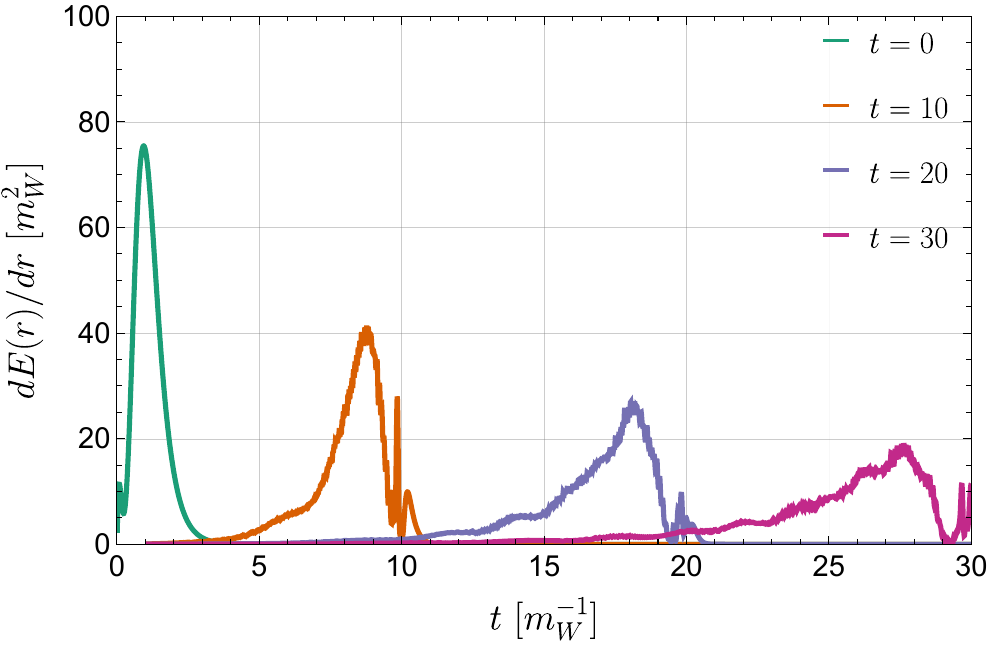}
    \end{minipage}

    \caption{Evolution of field profiles during sphaleron decay for Case II (large displacement, no initial kinetic energy). The arrangement and times shown are the same as in Fig.~\ref{fig:sphaleron_funcs_1}.}
    \label{fig:sphaleron_funcs_2}
\end{figure*}

\begin{figure*}[t!]
    \centering
    \begin{minipage}{0.48\linewidth}
        \centering
        \includegraphics[width=\linewidth]{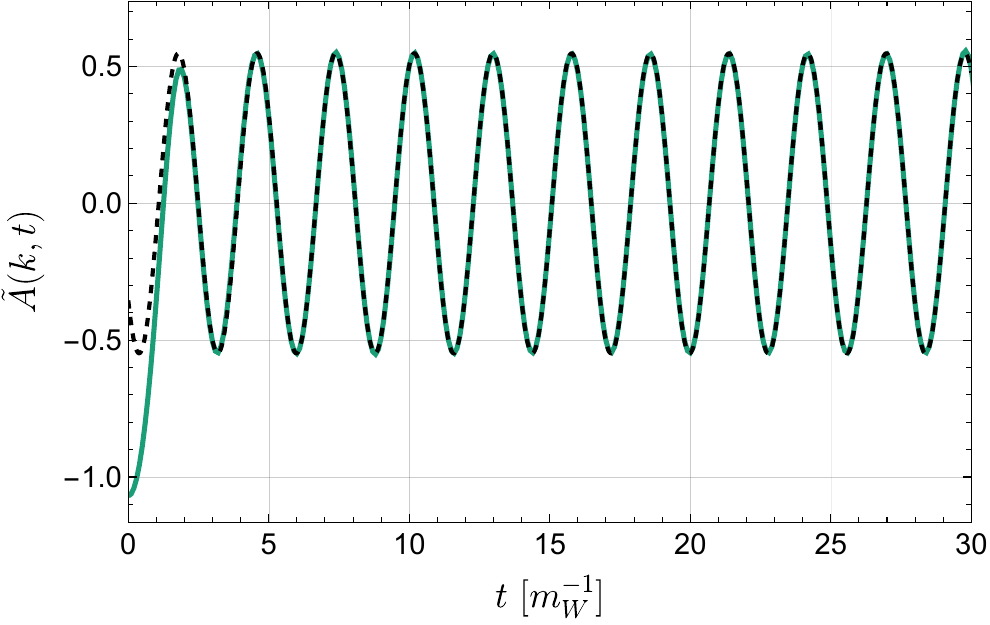}
    \end{minipage}
    \hfill
    \begin{minipage}{0.48\linewidth}
        \centering
        \includegraphics[width=\linewidth]{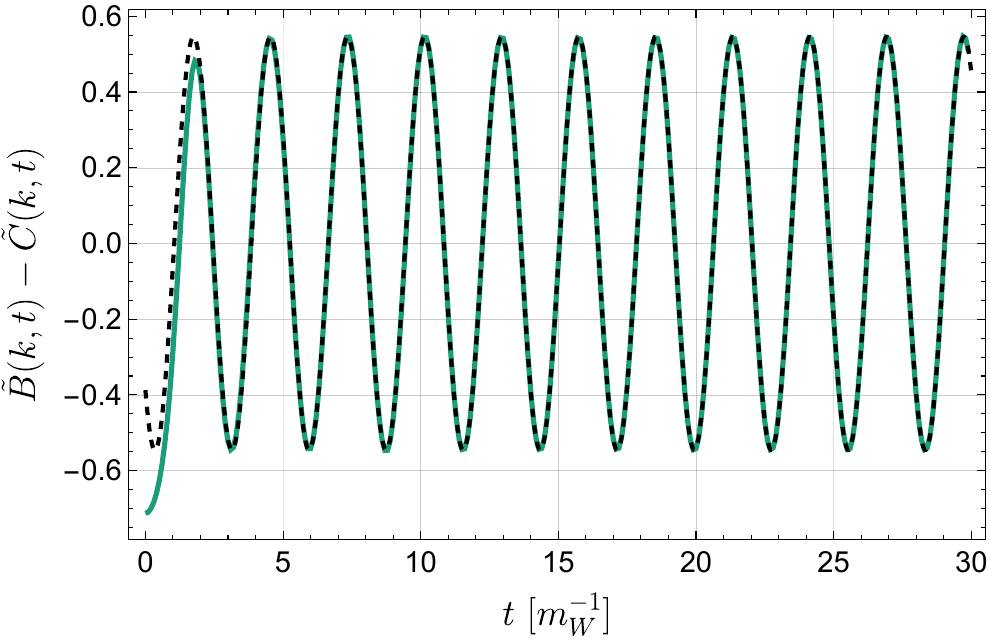}
    \end{minipage}
    
    \vspace{1em}
    
    \begin{minipage}{0.48\linewidth}
        \centering
        \includegraphics[width=\linewidth]{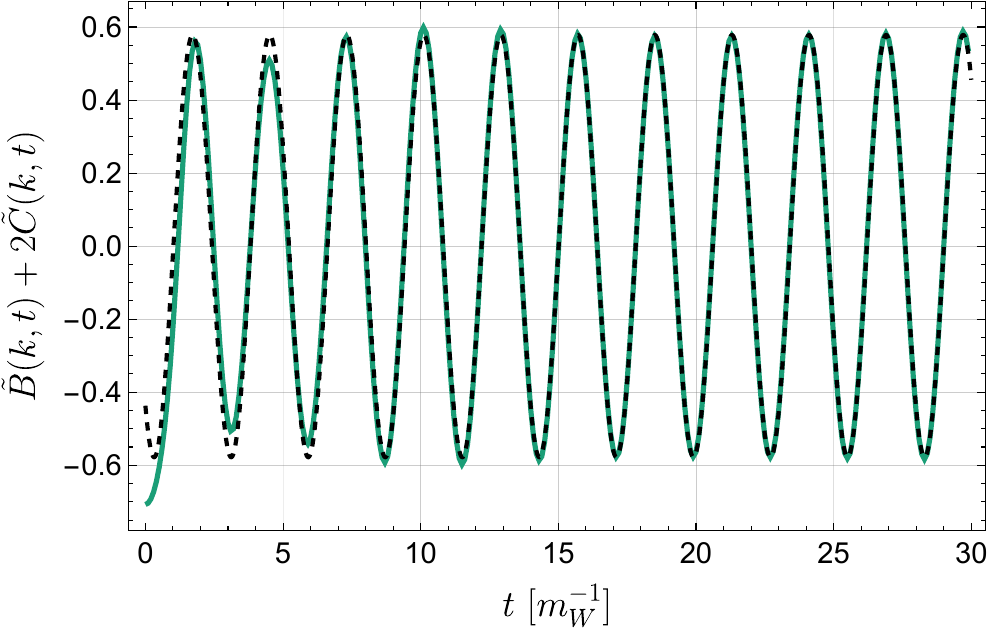}
    \end{minipage}
    \hfill
    \begin{minipage}{0.48\linewidth}
        \centering
        \includegraphics[width=\linewidth]{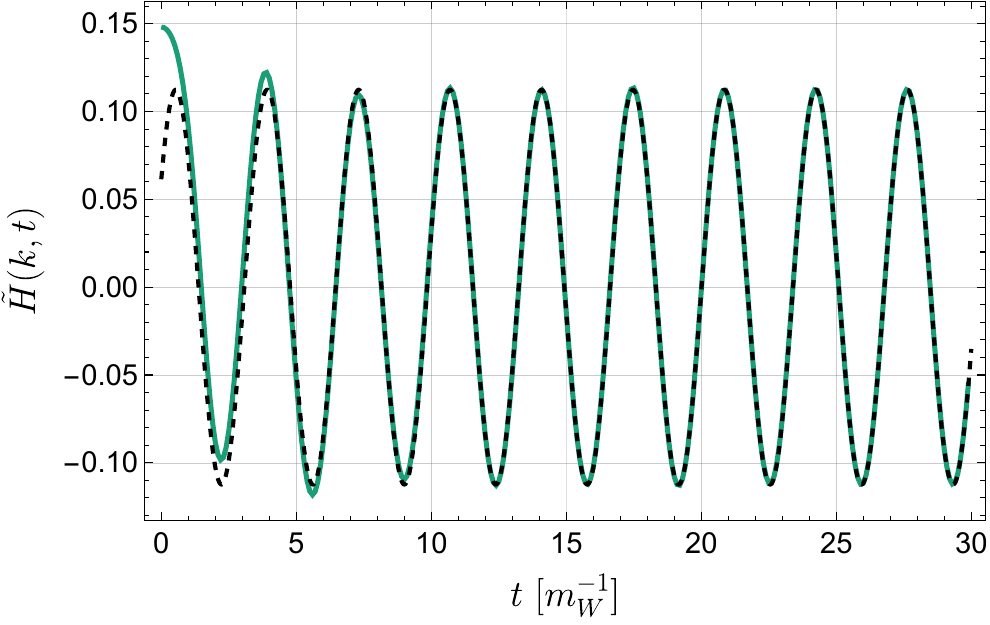}
    \end{minipage}

    \caption{Time evolution of spectral amplitudes at fixed momentum $k = 2m_W$ for Case II, arranged as in Fig.~\ref{fig:sphaleron_fourier_1}. }
    \label{fig:sphaleron_fourier_2}
\end{figure*}

To investigate how the decay characteristics depend on the magnitude of the initial perturbation, we now consider a larger displacement along the unstable direction. Specifically, we set:
\begin{align}
    \label{eq:case2_init1}
    f_B(r, 0) &= \sqrt{0.1} \times r\, \eta_B(r) \, , \\
    \label{eq:case2_init2}
    f_C(r, 0) &= \sqrt{0.1} \times \sqrt{2} \, \eta_C(r) \, , \\
    \label{eq:case2_init3}
    H(r, 0) &= \sqrt{0.1} \times \frac{1}{\sqrt{2} m_W}  \, \eta_H(r) \, , 
\end{align}
with the same normalization condition as in Case I and zero initial velocities. The coefficient $\sqrt{0.1} \simeq 0.316$ represents a significantly larger perturbation.

Figure~\ref{fig:sphaleron_funcs_2} shows the field evolution for this case. Compared to Case I, the decay proceeds more slowly and with smaller amplitude oscillations. The energy density distribution (bottom-right panel) shows a faster outward propagation of energy, consistent with the larger displacement. The spectral evolution, shown in Fig.~\ref{fig:sphaleron_fourier_2}, displays that the free-field oscillatory regime is reached faster. 

\begin{figure}[t!]
    \centering
    \includegraphics[width=1\linewidth]{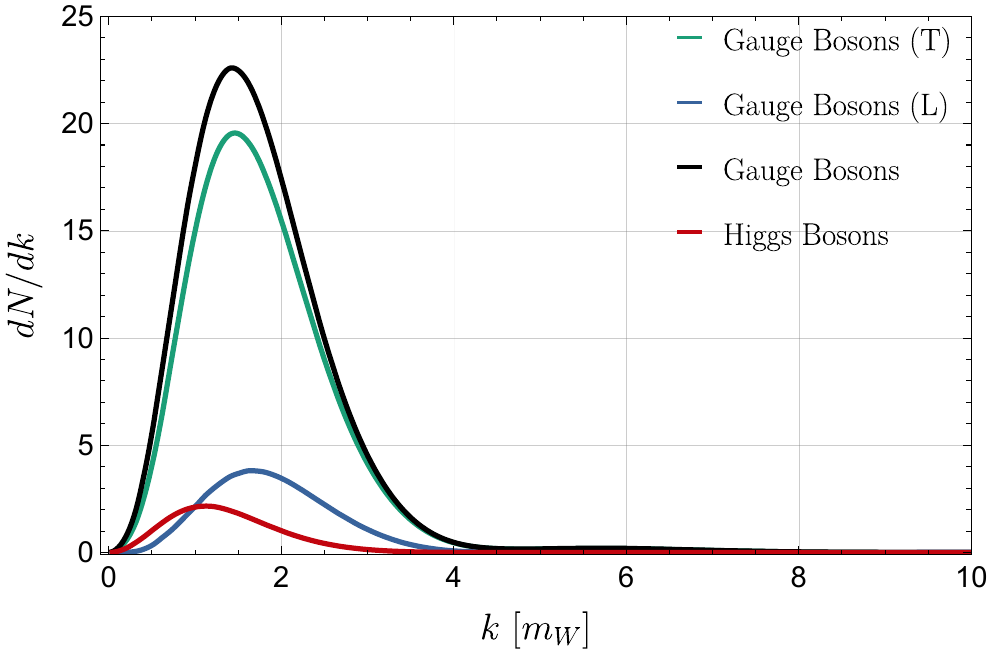}
    \caption{Particle multiplicity distributions $dN/dk$ as functions of momentum $k$ (in units of $m_W$) for Case II, with colors as in Fig.~\ref{fig:multiplicity1}. Compared to Case I, the distributions show lower overall multiplicities. The relative proportions of different particle species remain similar, with transverse gauge bosons still dominating the spectrum.}
    \label{fig:multiplicity2}
\end{figure}

The resulting particle multiplicity distributions, shown in Fig.~\ref{fig:multiplicity2}, maintain the same general characteristics as in Case I, but with lower overall multiplicities. The total multiplicities decrease to 35.1 for transverse gauge bosons, 6.9 for longitudinal gauge bosons (for a total of 42 gauge bosons), and 3.3 for Higgs bosons. We note that in this scenario the total multiplicity is significantly lower because the initial displacement along the unstable direction is substantially larger.

\subsection{Case III: No Displacement with Initial Kinetic Energy}
\label{subsec:case3}
\begin{figure*}[t!]
    \centering
    \begin{minipage}{0.48\linewidth}
        \centering
        \includegraphics[width=\linewidth]{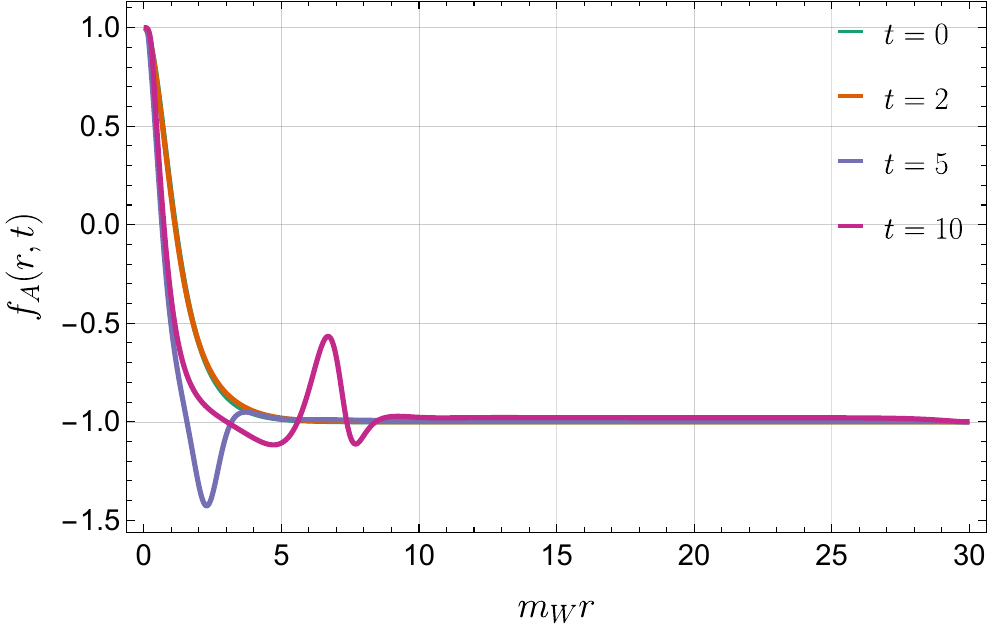}
    \end{minipage}
    \hfill
    \begin{minipage}{0.48\linewidth}
        \centering
        \includegraphics[width=\linewidth]{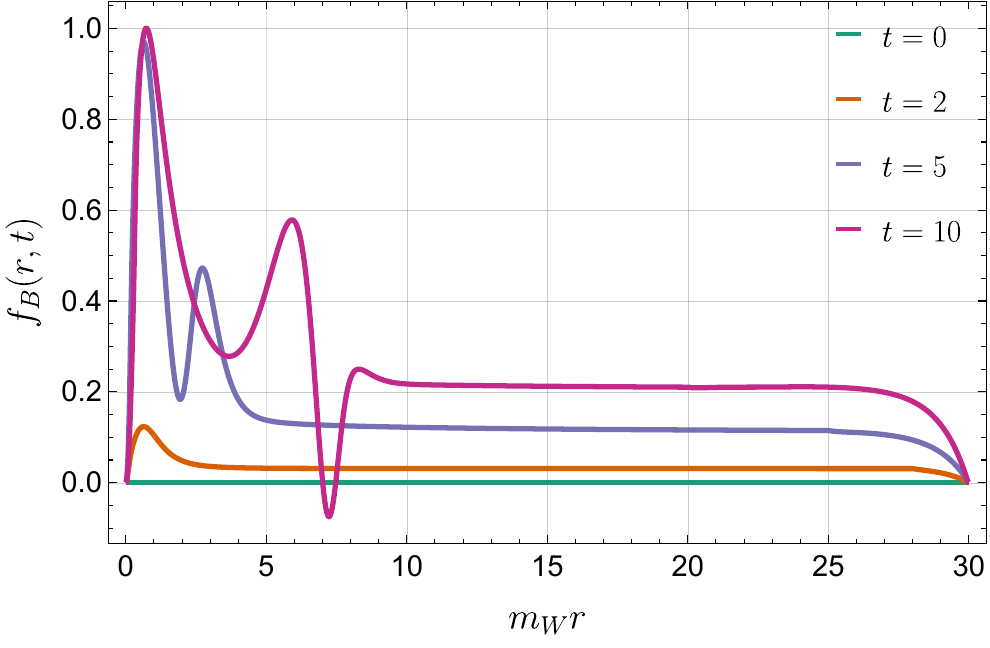}
    \end{minipage}
    
    \vspace{1em}
    
    \begin{minipage}{0.48\linewidth}
        \centering
        \includegraphics[width=\linewidth]{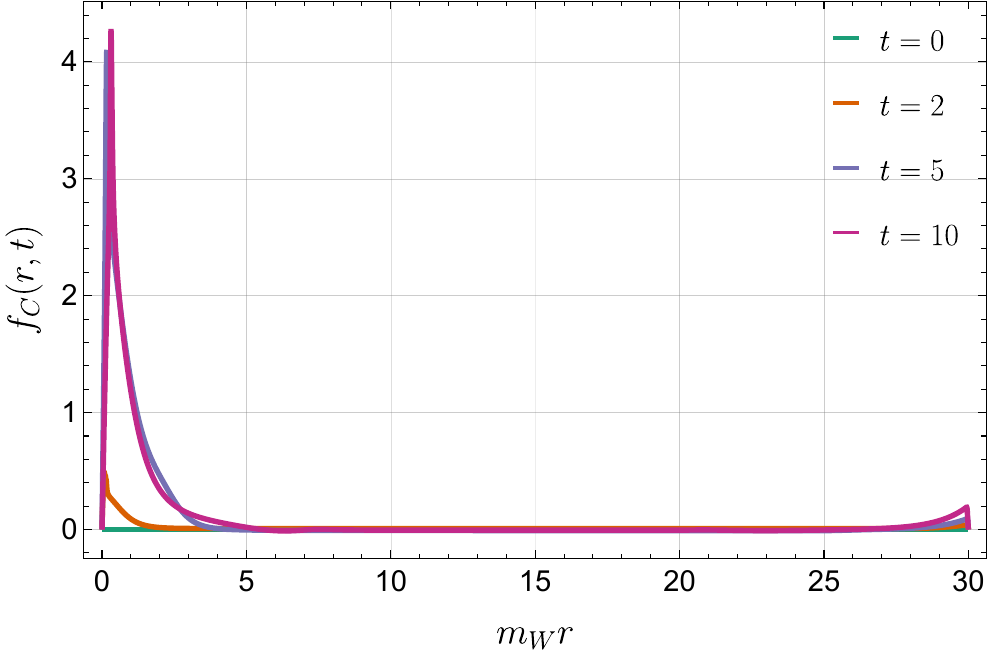}
    \end{minipage}
    \hfill
    \begin{minipage}{0.48\linewidth}
        \centering
        \includegraphics[width=\linewidth]{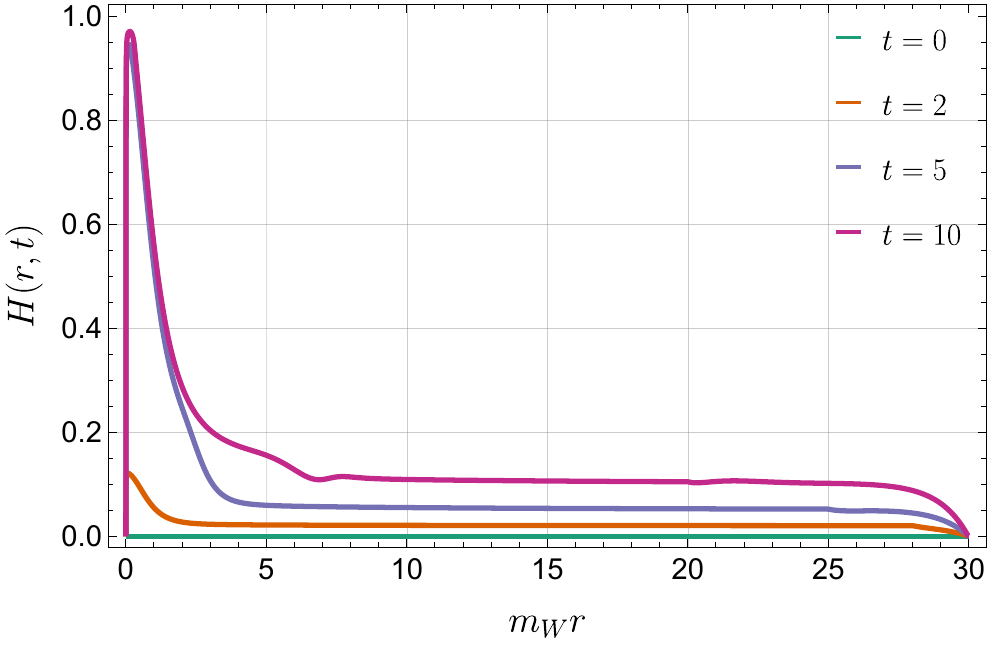}
    \end{minipage}
    
    \vspace{1em}
    
    \begin{minipage}{0.48\linewidth}
        \centering
        \includegraphics[width=\linewidth]{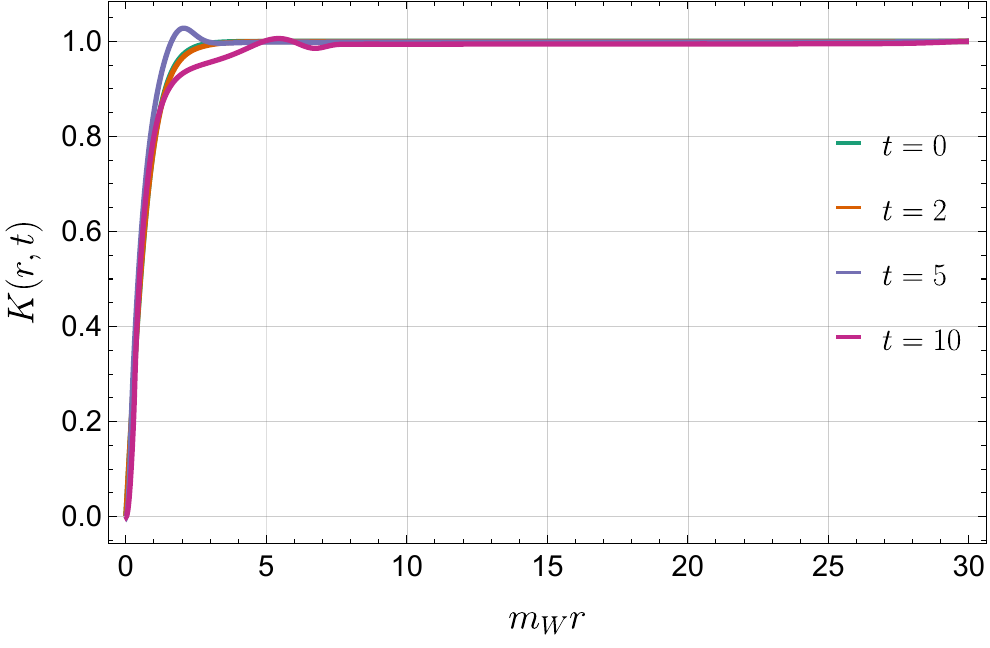}
    \end{minipage}
    \hfill
    \begin{minipage}{0.48\linewidth}
        \centering
        \includegraphics[width=\linewidth]{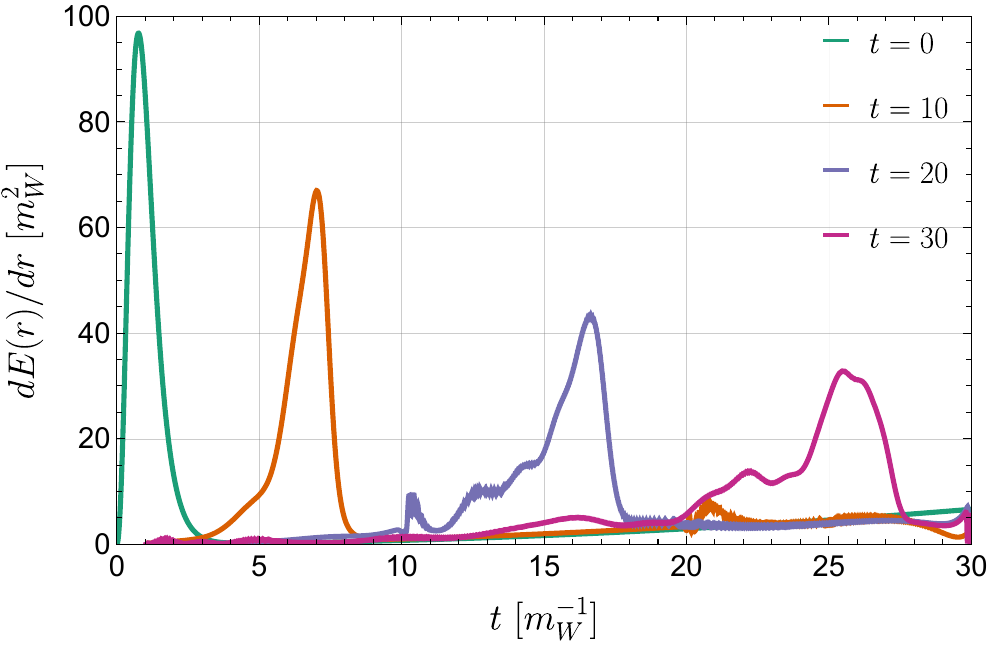}
    \end{minipage}

    \caption{Evolution of field profiles during sphaleron decay for Case III (no displacement with initial kinetic energy). The arrangement and times shown are the same as in Fig.~\ref{fig:sphaleron_funcs_1}.}
    \label{fig:sphaleron_funcs_3}
\end{figure*}

\begin{figure*}[t!]
    \centering
    \begin{minipage}{0.48\linewidth}
        \centering
        \includegraphics[width=\linewidth]{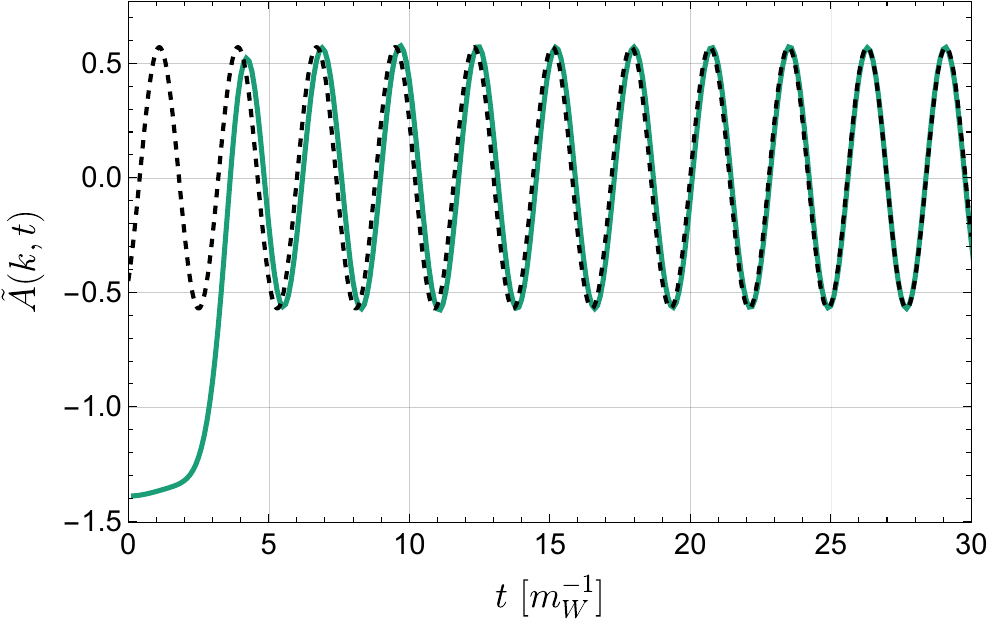}
    \end{minipage}
    \hfill
    \begin{minipage}{0.48\linewidth}
        \centering
        \includegraphics[width=\linewidth]{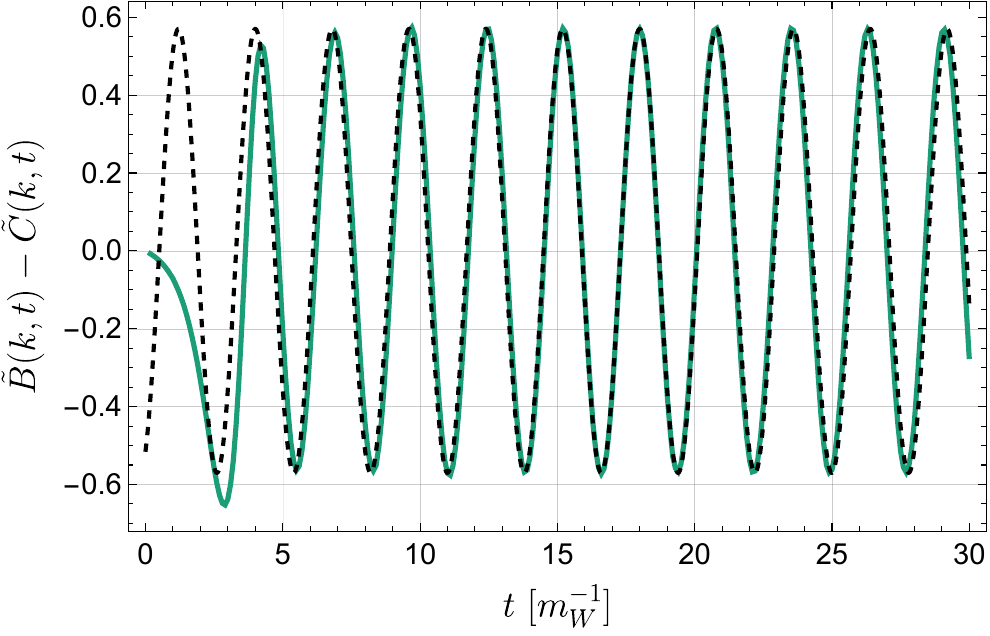}
    \end{minipage}
    
    \vspace{1em}
    
    \begin{minipage}{0.48\linewidth}
        \centering
        \includegraphics[width=\linewidth]{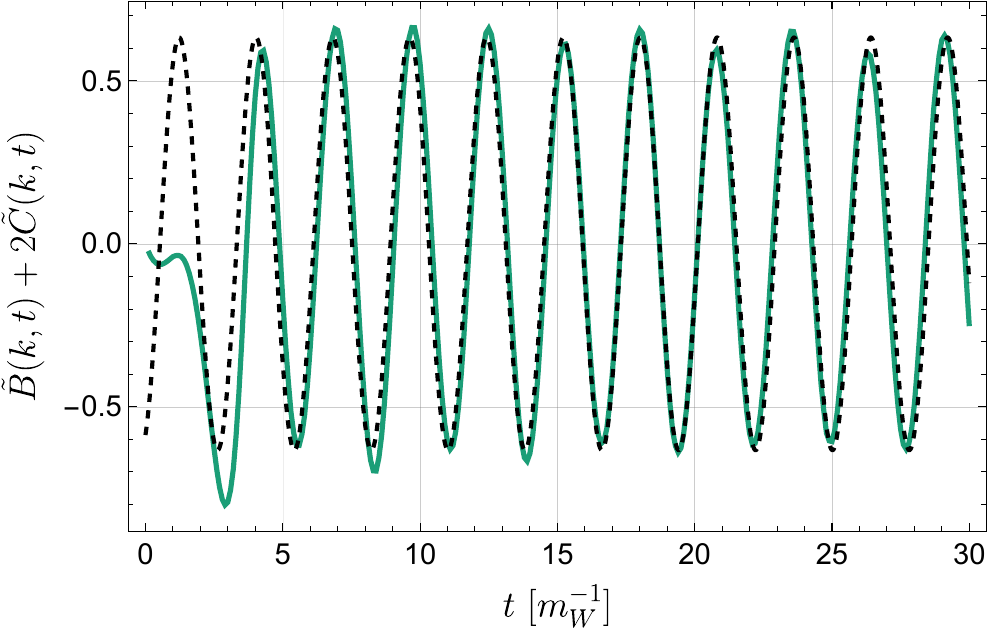}
    \end{minipage}
    \hfill
    \begin{minipage}{0.48\linewidth}
        \centering
        \includegraphics[width=\linewidth]{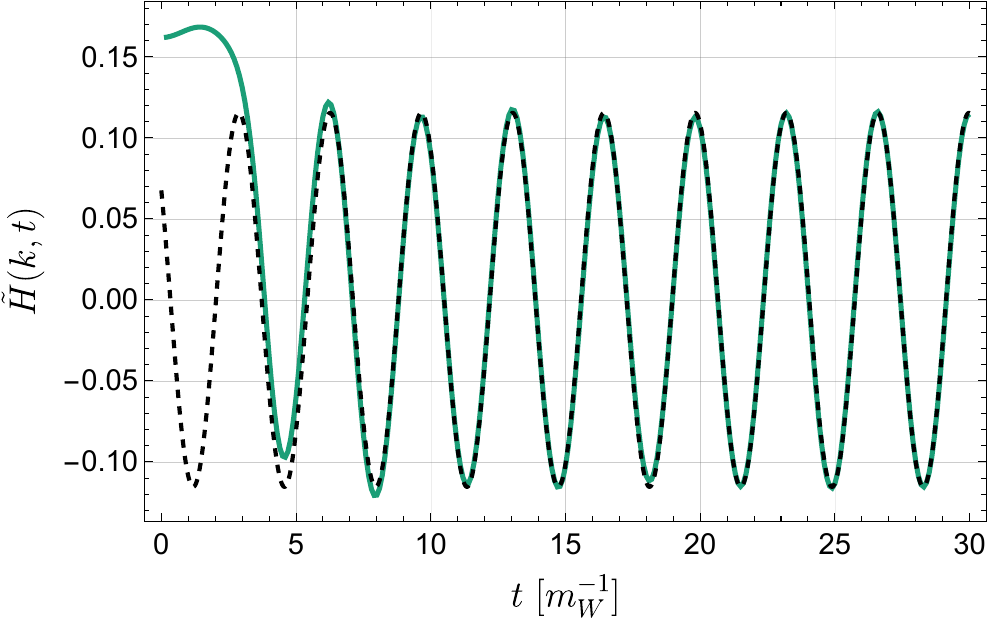}
    \end{minipage}
    \caption{Time evolution of spectral amplitudes at fixed momentum $k = 2m_W$ for Case III, arranged as in Fig.~\ref{fig:sphaleron_fourier_1}. }
    \label{fig:sphaleron_fourier_3}
\end{figure*}

We now investigate a qualitatively different decay scenario, where the sphaleron is not displaced from its equilibrium configuration but instead given an initial velocity along the unstable direction. This models situations where the decay is triggered by dynamical processes rather than by static field configurations. The initial conditions are:
\begin{align}
    \label{eq:case3_init1}
    f_B(r, 0) &= f_C(r, 0) = H(r, 0) = 0 \, . 
\end{align}
The velocity is chosen to be
\begin{align}
    \label{eq:case3_vel}
    \dot{f}_B(r,0) = \dot{f}_C(r,0) = \dot{H}(r,0) = 0.01 m_W \, .
\end{align}

\begin{figure}[t!]
    \centering
    \includegraphics[width=1\linewidth]{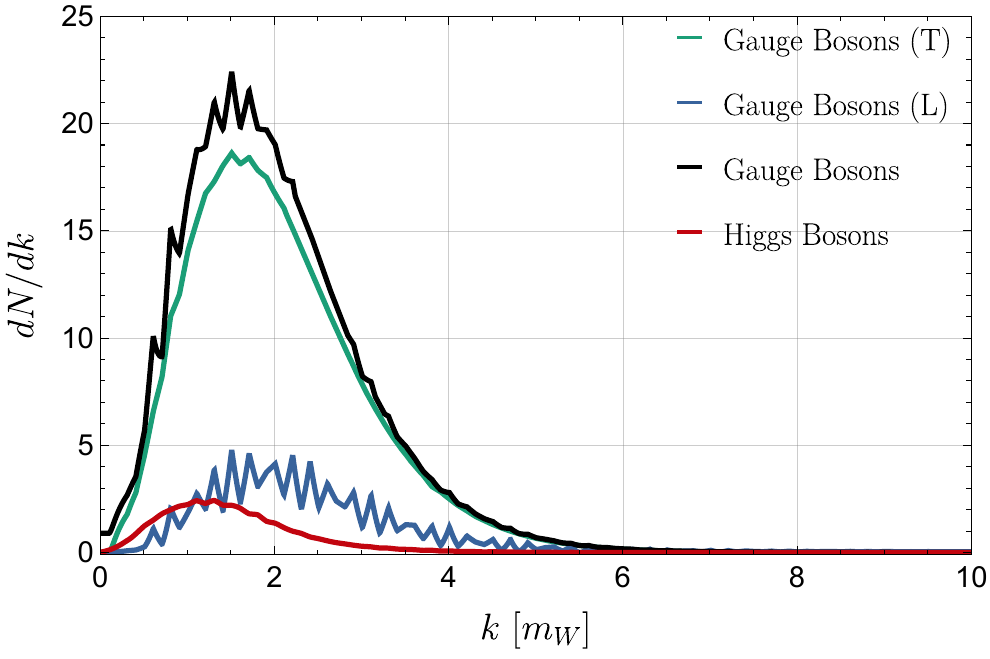}
    \caption{Particle multiplicity distributions $dN/dk$ as functions of momentum $k$ (in units of $m_W$) for Case III.}
    \label{fig:multiplicity3}
\end{figure}

Our simulations reveal that the overall decay process proceeds in a qualitatively similar manner to the cases with initial displacement (see Figures~\ref{fig:sphaleron_funcs_3}, \ref{fig:sphaleron_fourier_3} and \ref{fig:multiplicity3}). However, there are some notable differences in the early-time dynamics and in the final particle spectra. In particular, the initial phase exhibits more pronounced oscillatory behavior, as the energy is initially in the form of field momentum rather than potential energy. This leads to more complex mode mixing and a somewhat broader spectrum of produced particles.

The total particle multiplicities in this case are 41.8 for transverse gauge bosons, 8.6 for longitudinal gauge bosons (for a total of 50.4 gauge bosons), and 4.1 for Higgs bosons. The energy carried by Higgs bosons accounts for 7.1\% of the total, slightly lower than in the displacement-triggered scenarios.

The spectral distributions in this case show more fine-scale structure and are generally noisier than in Cases I and II. This is a consequence of the more complex initial dynamics, which excite a broader range of modes. These features could potentially be resolved with higher numerical resolution, but the overall physics conclusions remain robust.

\subsection{Case IV: Small Displacement with Initial Kinetic Energy}
\label{subsec:case4}
\begin{figure*}[t!]
    \centering
    \begin{minipage}{0.48\linewidth}
        \centering
        \includegraphics[width=\linewidth]{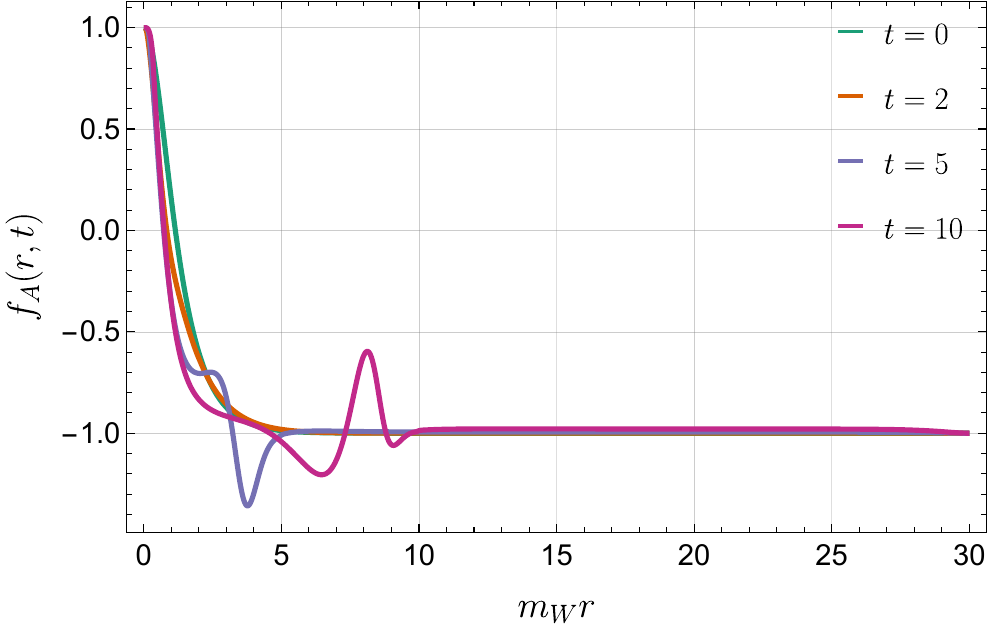}
    \end{minipage}
    \hfill
    \begin{minipage}{0.48\linewidth}
        \centering
        \includegraphics[width=\linewidth]{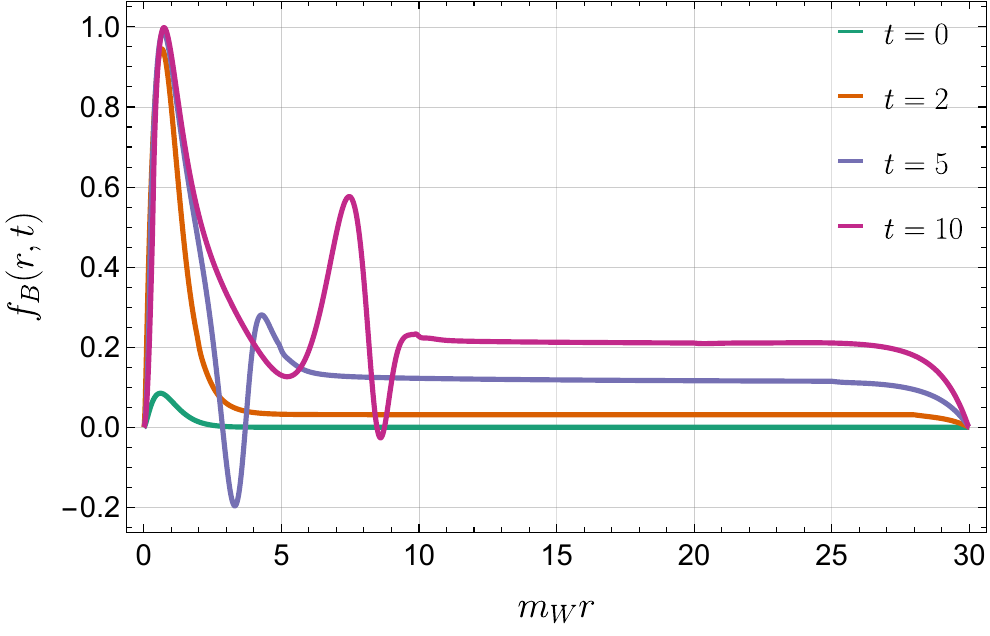}
    \end{minipage}
    
    \vspace{1em}
    
    \begin{minipage}{0.48\linewidth}
        \centering
        \includegraphics[width=\linewidth]{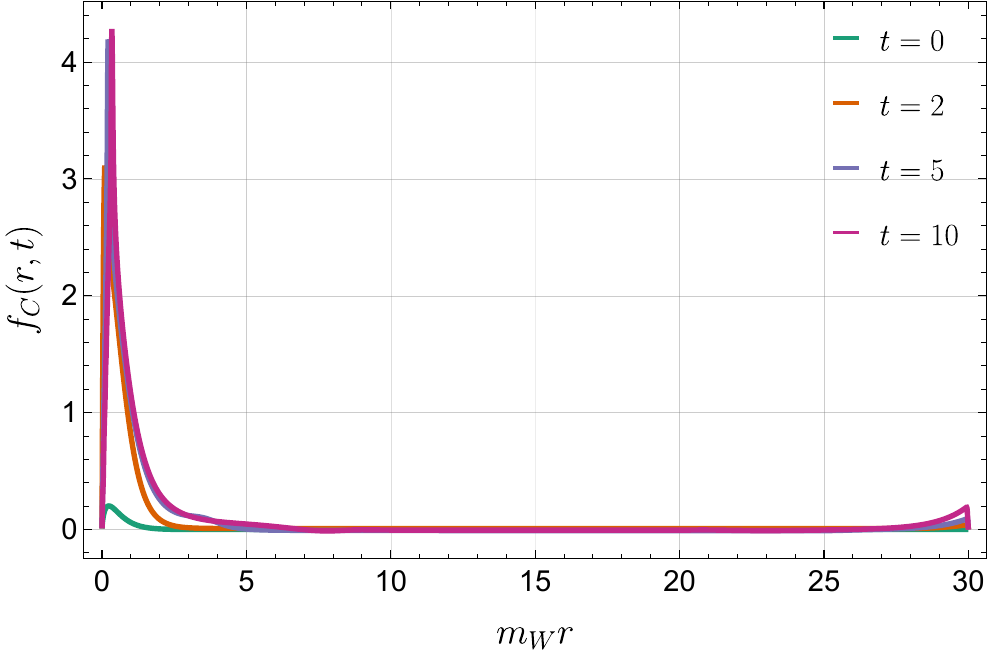}
    \end{minipage}
    \hfill
    \begin{minipage}{0.48\linewidth}
        \centering
        \includegraphics[width=\linewidth]{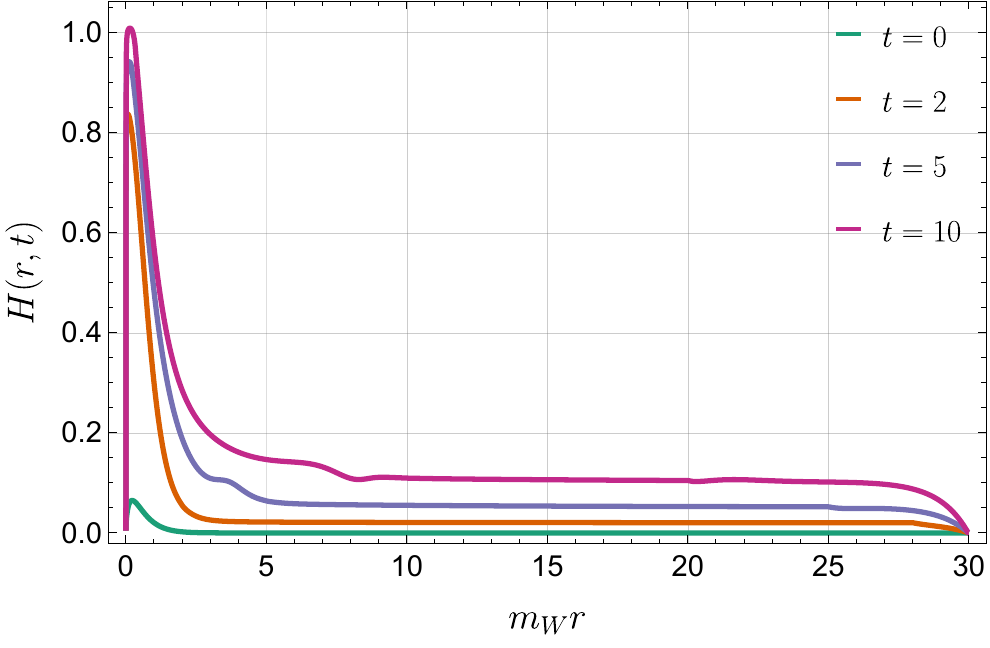}
    \end{minipage}
    
    \vspace{1em}
    
    \begin{minipage}{0.48\linewidth}
        \centering
        \includegraphics[width=\linewidth]{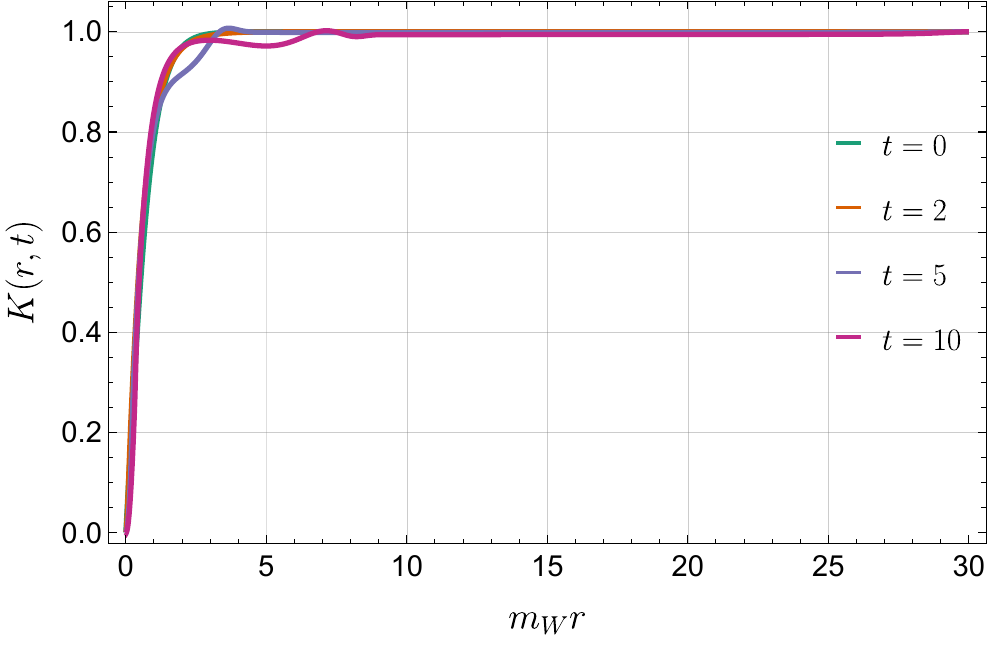}
    \end{minipage}
    \hfill
    \begin{minipage}{0.48\linewidth}
        \centering
        \includegraphics[width=\linewidth]{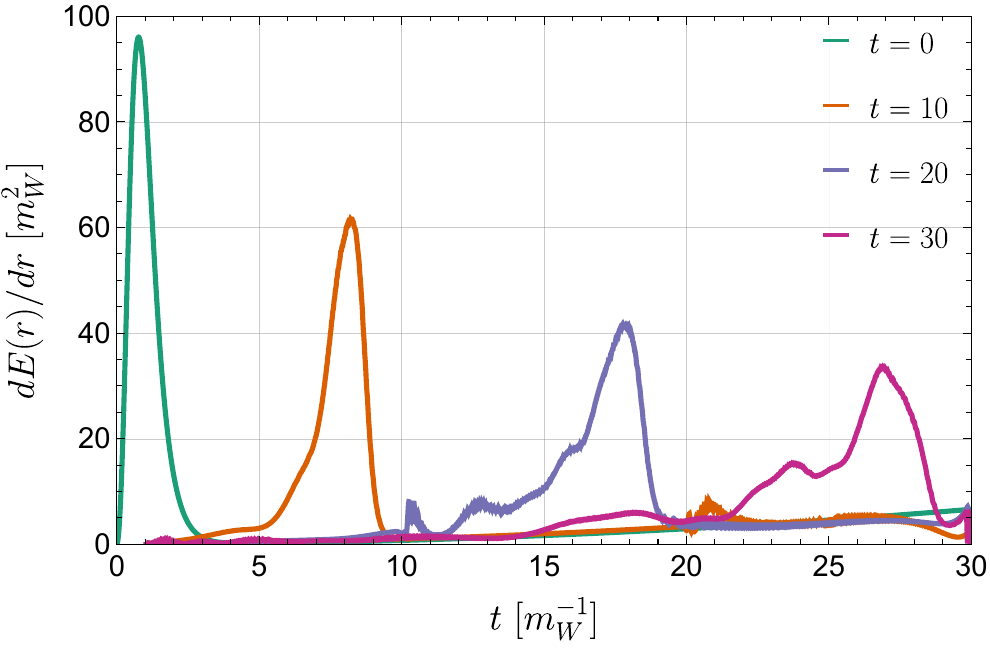}
    \end{minipage}

    \caption{Evolution of field profiles during sphaleron decay for Case IV (small displacement with initial kinetic energy). The arrangement and times shown are the same as in Fig.~\ref{fig:sphaleron_funcs_1}.}
    \label{fig:sphaleron_funcs_4}
\end{figure*}

\begin{figure*}[t!]
    \centering
    \begin{minipage}{0.48\linewidth}
        \centering
        \includegraphics[width=\linewidth]{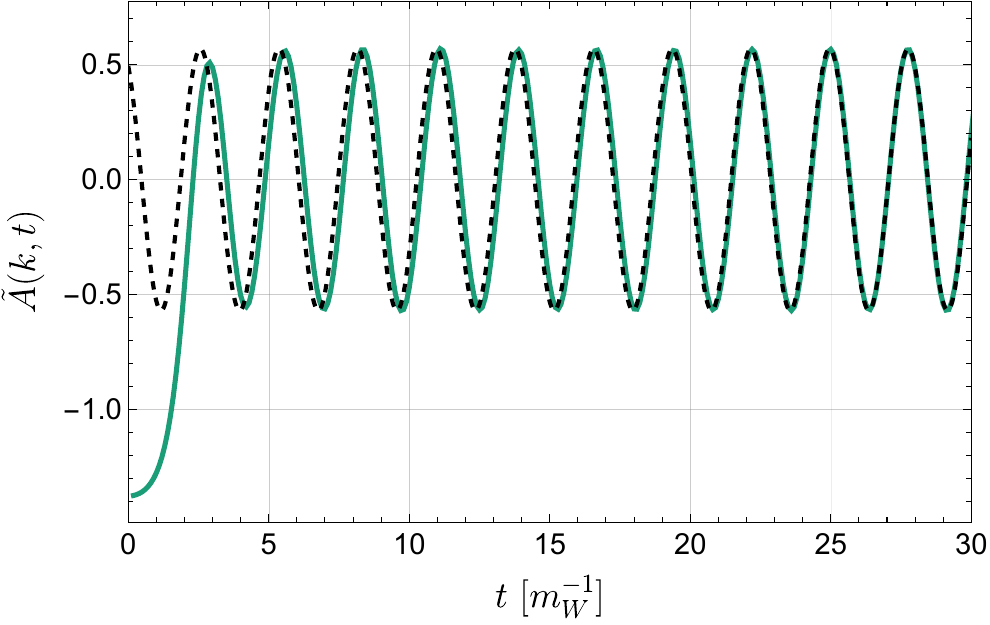}
    \end{minipage}
    \hfill
    \begin{minipage}{0.48\linewidth}
        \centering
        \includegraphics[width=\linewidth]{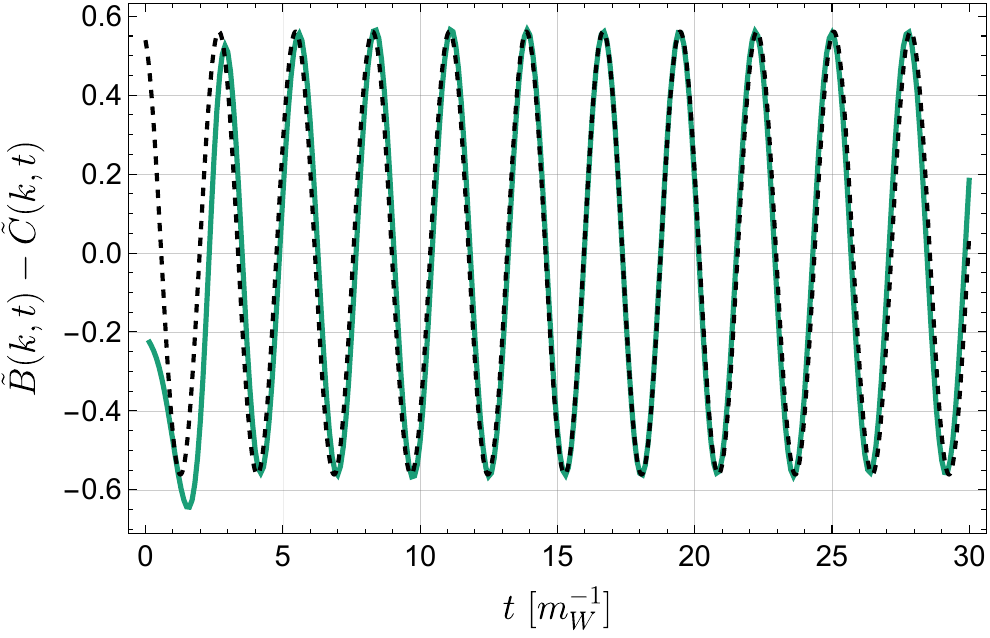}
    \end{minipage}
    
    \vspace{1em}
    
    \begin{minipage}{0.48\linewidth}
        \centering
        \includegraphics[width=\linewidth]{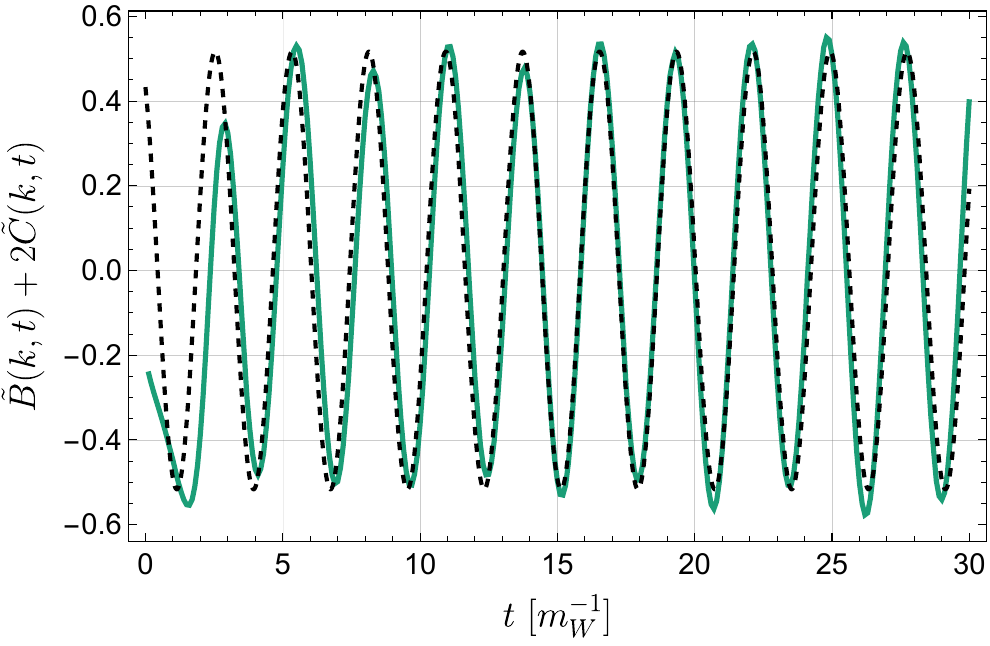}
    \end{minipage}
    \hfill
    \begin{minipage}{0.48\linewidth}
        \centering
        \includegraphics[width=\linewidth]{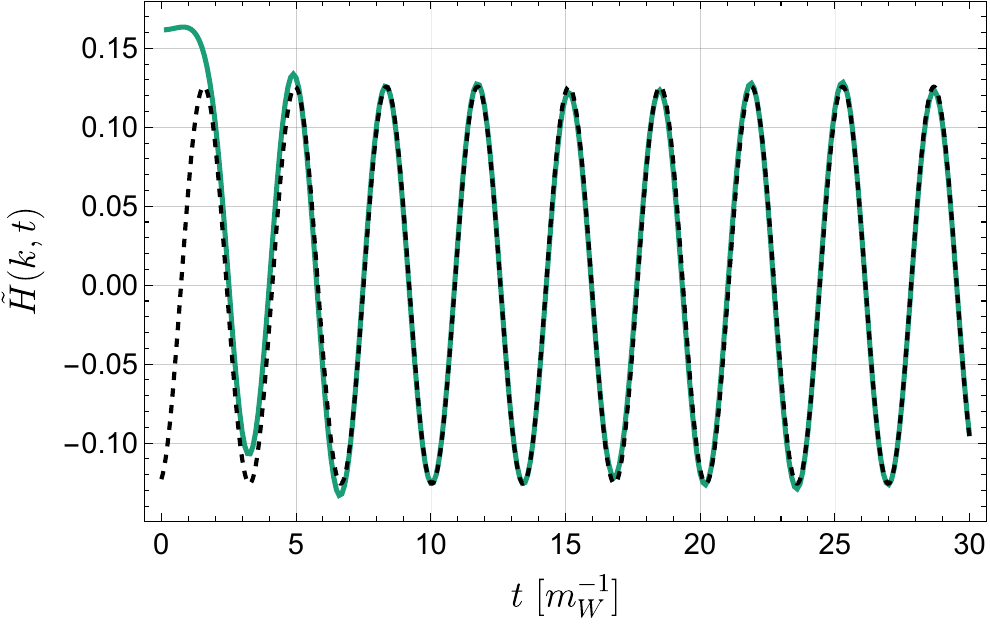}
    \end{minipage}
    \caption{Time evolution of spectral amplitudes at fixed momentum $k = 2m_W$ for Case IV, arranged as in Fig.~\ref{fig:sphaleron_fourier_1}. }
    \label{fig:sphaleron_fourier_4}
\end{figure*}

\begin{figure}[t!]
    \centering
    \includegraphics[width=1\linewidth]{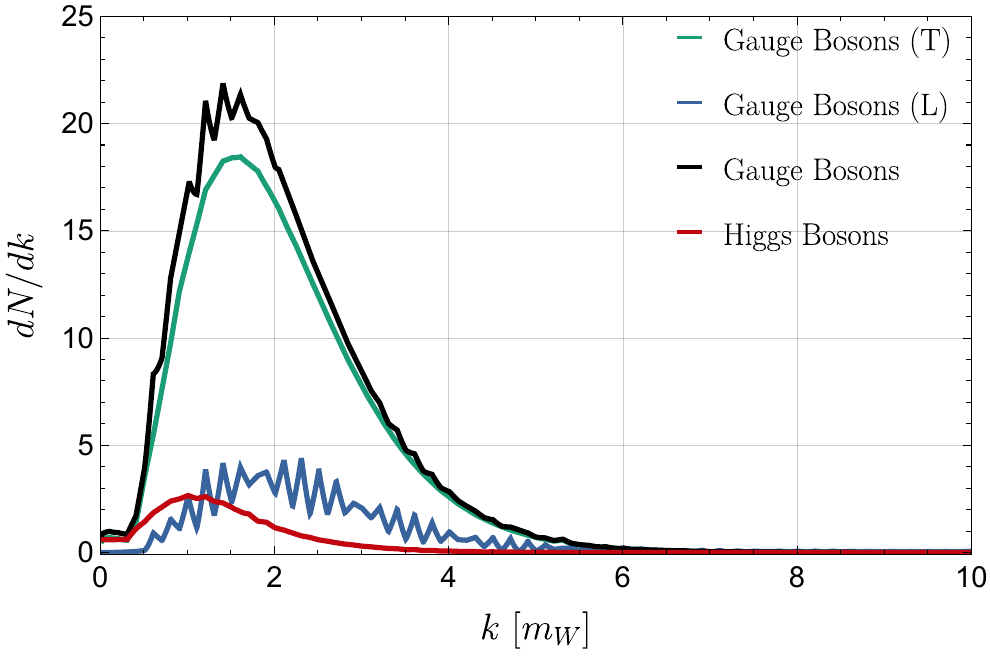}
    \caption{Particle multiplicity distributions $dN/dk$ as functions of momentum $k$ (in units of $m_W$) for Case IV.}
    \label{fig:multiplicity4}
\end{figure}

Our final scenario combines both displacement and initial velocity along the unstable direction, representing the most general decay configuration. We set:
\begin{align}
    \label{eq:case4_init1}
    f_B(r, 0) &= 0.1 \times r\, \eta_B(r) \, , \\
    \label{eq:case4_init2}
    f_C(r, 0) &= 0.1 \times \sqrt{2} \, \eta_C(r) \, , \\
    \label{eq:case4_init3}
    H(r, 0) &= 0.1 \times \frac{1}{\sqrt{2} m_W}  \, \eta_H(r) \, , 
\end{align}
and the initial velocity
\begin{align}
    \label{eq:case4_vel}
    \dot{f}_B(r,0) = \dot{f}_C(r,0) = \dot{H}(r,0) = 0.01 m_W \, .
\end{align}

The decay dynamics in this general case combines features from both the displacement-triggered and velocity-triggered scenarios (see Figure~\ref{fig:sphaleron_funcs_4} and Figure~\ref{fig:sphaleron_fourier_4}). The early evolution shows a mix of exponential growth (from the displacement component) and oscillatory behavior (from the velocity component). However, after the initial transient phase, the system evolves toward a similar final state as in the previous cases.

In this scenario, we find that the total particle multiplicities in are 40.9 for transverse gauge bosons, 8.4 for longitudinal gauge bosons (for a total of 49.3 gauge bosons), and 4.4 for Higgs bosons. The energy carried by Higgs bosons accounts for 7.4\% of the total, slightly lower than in the displacement-triggered scenarios.

\section{Conclusions}
\label{sec:conclusions}
In this paper, we have presented a comprehensive investigation of sphaleron decay dynamics in the $\mathrm{SU}(2)_W$ theory, combining analytical characterization with high-resolution numerical simulations. Our work establishes a detailed understanding of how these topologically non-trivial configurations evolve and produce particles, with implications for both early universe cosmology and high-energy collider physics.

We began by formulating the spherically symmetric ansatz for the $\mathrm{SU}(2)_W$ system, deriving the classical field equations that govern the sphaleron and its evolution. Our stability analysis confirmed the existence of precisely one unstable mode with eigenvalue $\omega_{-}^2 \simeq -2.7m_W^2$, in agreement with previous studies but now calculated with current Standard Model parameters ($m_H = 125.1$ GeV, $m_W = 80.4$ GeV)~\cite{MatchevVerner}. This unstable mode defines the preferred direction for sphaleron decay and the resulting topological transition between neighboring vacua.

We developed a systematic approach for analyzing the asymptotic behavior of the fields after the sphaleron decay, introducing a spherical wave decomposition that relates the classical field oscillations to particle production. This framework allowed us to quantify the energy and multiplicity distributions of gauge and Higgs bosons produced during the decay process. We found that the sphaleron decay process is remarkably universal across different initial conditions. Whether triggered by field displacements or initial velocities along the unstable direction, the decay follows similar patterns and yields comparable particle distributions, suggesting that the intrinsic properties of the sphaleron largely determine the decay outcomes.

The particle production is dominated by transverse gauge bosons, which consistently account for approximately 75\% - 80\% of the total energy and multiplicity. This preference can be traced to the direct connection between the transverse polarization state and the topological structure of the electroweak theory. The spectral distributions of produced particles peak at momenta $k \sim 1 - 1.5 m_W$, corresponding to wavelengths matching the characteristic size of the sphaleron. This establishes a direct link between the spatial structure of non-perturbative field configurations and the energy scale of resulting particles.

Higgs bosons constitute only a small fraction (7-8\%) of the decay products despite their essential role in defining the topological structure. This asymmetry reflects the specifics of how gauge and Higgs fields contribute to the unstable mode of the sphaleron. While increasing the initial perturbation magnitude enhances the overall particle production, the relative distribution among different species remains almost constant, suggesting a robust pattern determined by fundamental aspects of the theory rather than by specific decay conditions.

For high-energy collider physics, our findings suggest that if sphaleron-like transitions were to become accessible at future accelerators, they would manifest primarily through the production of multiple transverse $W$ and $Z$ bosons with a characteristic momentum distribution. This provides a potential experimental signature distinct from perturbative processes.

Several important directions remain for future investigation. First, extending our analysis to include the full electroweak theory with physical $\sin^2\theta_W \simeq 0.23$ would provide more precise predictions for realistic scenarios. Second, incorporating fermions would enable direct calculation of the baryon and lepton number violation during the sphaleron transition. Third, studying thermal effects would clarify how sphaleron transitions proceed in the hot early universe environment.

Finally, while our classical field theory approach captures the essential non-perturbative dynamics, a full quantum treatment would be necessary to address phenomena such as quantum tunneling between topological sectors at energies below the sphaleron barrier. Developing methods to bridge classical and quantum descriptions of topological transitions remains an important challenge for the field.

\begin{acknowledgments}
We thank Pierre Ramond for useful discussions. This work was supported in part by the U.S. Department of Energy award number DE-SC0022148. The work of KTM is supported in part by the Shelby Endowment for Distinguished Faculty at the University of Alabama. The work of S.V. was supported in part by DOE grant grant DE-SC0022148 at the University of Florida.
\end{acknowledgments}

\bibliography{references}

\end{document}